\documentclass[reqno,12pt]{article}
\usepackage{amsmath,amsfonts,amssymb,amsthm,amstext,amscd,eucal}
\usepackage[all]{xy}

\makeatletter \@addtoreset{equation}{section}

\makeatletter\renewcommand\section{\@startsection {section}{1}{\z@}%
                                   {-3.5ex \@plus -1ex \@minus -.2ex}
                                   {2.3ex \@plus.2ex}%
                                   {\normalfont\large\bfseries}}
\renewcommand\subsection{\@startsection{subsection}{2}{\z@}%
                                     {-3.25ex\@plus -1ex \@minus -.2ex}%
                                     {1.5ex \@plus .2ex}%
                                     {\normalfont\bfseries}}

\newcommand{\be}{\begin{equation}}
\newcommand{\ee}{\end{equation}}
\newcommand{\beq}{\begin{eqnarray}}
\newcommand{\eeq}{\end{eqnarray}}
\newcommand{\bea}{\begin{eqnarray}}
\newcommand{\eea}{\end{eqnarray}}

\def\[{\left [}
\def\]{\right ]}
\def\({\left (}
\def\){\right )}

\def\CN{{\cal N}}

\def\r2{\sqrt{2}}

\newcommand{\eref}[1]{(\ref{#1})}

\newcommand{\BZ}{{\bf Z}}

\newcommand{\tr}[1]{{\rm tr}[{#1}]}

\newcommand{\Polya}{P\'olya}
\newcommand{\AdS}[1]{{\rm AdS}_{#1}}

\def\CN{{\cal N}}

\newcommand{\ads}[1]{{\rm AdS}_{#1}}

\newcommand{\bbibitem}[1]{\bibitem{#1}\marginpar{#1}}

\def\Label#1{\label{#1}%
  \smash{\hbox to0pt{\raise1ex\hbox{\tiny[#1]}\hss}}}
\def\noLabels{\let\Label=\label}
\def\nobbibitem{\let\bbibitem=\bibitem}

\begin{document}
\noLabels 
\nobbibitem 

\begin{titlepage}

\begin{flushright}
{\small
UPR-T-1185,
ITFA-2007-33, \\
DCPT-07/37,
UCB-PTH-07/13}
\end{flushright}

\vfil\

\begin{center}

{\Large{\bf Entropy of near-extremal black holes in ${\rm AdS}_5$}}
\vfil
\vspace{3mm}

{\bf
Vijay Balasubramanian\footnote{e-mail: vijay@physics.upenn.edu}$^{,a}$,
Jan de Boer\footnote{e-mail: jdeboer@science.uva.nl}$^{,b}$,
Vishnu Jejjala\footnote{e-mail: vishnu.jejjala@durham.ac.uk}$^{,c}$,
and Joan Sim\'on\footnote{e-mail: JSimonSoler@lbl.gov}$^{,d}$
}
\\

\vspace{8mm}

\bigskip\medskip
\centerline{$^a$ \it David Rittenhouse Laboratories, University of Pennsylvania, Philadelphia, PA 19104, USA}
\smallskip
\centerline{$^b$ \it Instituut voor Theoretische Fysica, Valckenierstraat 65, 1018XE Amsterdam, The Netherlands}
\smallskip
\centerline{$^c$ \it Department of Mathematical Sciences, University of Durham, South Road, Durham DH1 3LE, UK}
\smallskip
\centerline{$^d$ \it Dept of Physics \& Theoretical Physics Group, LBNL,University of California, Berkeley, CA 94720, USA}

\vfil

\end{center}
\setcounter{footnote}{0}
\begin{abstract}
\noindent
We construct the microstates of near-extremal black holes in $\ads{5} \times S^5$ as gases of defects distributed in heavy BPS operators in the dual $SU(N)$ Yang--Mills theory.
These defects describe open strings on spherical D3-branes in the $S^5$, and we show that they dominate the entropy by directly enumerating them and comparing the results with a partition sum calculation.
We display new decoupling limits in which the field theory of the lightest open strings on the D-branes becomes dual to a near-horizon region of the black hole geometry.
In the single-charge black hole we find evidence for an infrared duality between $SU(N)$ Yang--Mills theories that exchanges the rank of the gauge group with an R-charge.
In the two-charge case (where pairs of branes intersect on a line), the decoupled geometry includes an $\ads{3}$ factor with a two-dimensional CFT dual.
The degeneracy in this CFT accounts for the black hole entropy.
In the three-charge case (where triples of branes intersect at a point), the decoupled geometry contains an $\ads{2}$ factor.
Below a certain critical mass, the two-charge system displays solutions with naked timelike singularities even though they do not violate a BPS bound.
We suggest a string theoretic resolution of these singularities.
\end{abstract}
\vspace{0.5in}

\end{titlepage}
\renewcommand{\baselinestretch}{1.05}  
\tableofcontents

\newpage

\section{Introduction}

There is a family of asymptotically $\ads{5} \times S^5$ black holes charged under three $U(1)$ fields \cite{cveticetal}.
The source for the geometry is a condensate of spherical D-branes (giant gravitons) localized on the $S^5$ and carrying angular momentum in three orthogonal planes \cite{superstar}.
The BPS black holes in this class (superstars) have a vanishing horizon, but as energy is added at fixed charge they eventually develop a finite horizon.
While the horizon size is a continuous function of the energy above extremality for the singly charged black hole, it is a discontinuous function of energy for the multiply charged black holes.

It has been shown in the dual $SU(N)$ Yang--Mills theory that the D-brane configurations giving rise to the entropy of the extremal single-charge black hole all lie very close to a certain {\it typical state} \cite{babel1,quantgrav}.
Here we treat the near-extremal black hole as a gas of defects distributed on the BPS operator describing a typical state \cite{BBNS,cjr,ber}.
The defects describe open strings on the giant gravitons in $\ads{5} \times S^5$ \cite{BHNL,davidemergence,Berensteinetal,BBFH,MelloKoch}, and we show that they dominate the entropy at weak coupling by directly enumerating them and comparing the results with a partition sum calculation.\footnote{A similar philosophy has been applied in the context of large supersymmetric $\ads{5}\times S^5$ black holes in \cite{sonner}, where quantization of the classical moduli space of probe giants (and generalizations thereof) in the black hole near horizon geometry accounts for their gravitational entropy.}
The degeneracy associated with the defects turns out to be independent of the string coupling and scales with $N$ in the same way as the black hole entropy when the horizon size is comparable to the AdS scale and hence thermodynamically stable.
However, there is a mismatch with the Bekenstein--Hawking entropy by factor of $O(1)$ (in terms of scaling in $N$) which depends on the distance from extremality.

We then display various decoupling limits where we send $\ell_s\rightarrow 0$ while keeping the masses of the open strings stretched between giant gravitons fixed.   In geometric terms the decoupling limits focus into the near-horizon region and onto some angles of the $S^5$ factor of the spacetimes.  In the one-charge case this results in a metric which looks like the near-horizon metric of a stack of giant gravitons that wrap an $S^3$ and that are smeared along a transversal two-cycle.
Exactly the same metric can also be obtained as the decoupling limit of a different geometry where the roles of the number of D3-branes and the number of giant gravitons are interchanged.  This provides evidence for the existence of the infrared duality proposed in \cite{vijayasad} under which the rank of the gauge group and the number of giant gravitons are interchanged.

A similar decoupling limit focuses on fixed angles in the $S^5$ factor of  the non-extremal two-charge $\ads{5}$ black hole.  This results in a geometry with an asymptotic $\ads{3}$ factor.   Using the asymptotic conformal symmetry of this factor, we measure a central charge from the geometry. Assuming that this is the central charge of a two-dimensional CFT dual, we find an expression for the statistical degeneracy of the CFT  that exactly matches the spacetime entropy  of the decoupled geometry.   The entropy of the complete two-charge $\ads{5}$ black hole can be recovered as an integral over the entropies of the decoupled geometries.  This suggests that the complete near-horizon geometry is dual to a product of the two-dimensional CFTs arising from our decoupling limit.  (The appearance of a two-dimensional CFT recalls the arguments of \cite{gubserheckman} who used a CFT associated to the intersection of giant gravitons to study the two-charge black hole entropy.)  Our formulation also explains a puzzle about the two-charge $\ads{5}$ black holes --- a horizon only develops above a certain critical energy above extremality.
In the decoupling limit, the solutions below the critical energy give rise to conical defects in the $\ads{3}$ while above the critical energy BTZ black hole factors appear.
As we will discuss, we expect that these conical defect metrics are resolved in string theory.

In the three-charge case, the decoupling limit results in a metric which contains an $\ads{2}$ factor.
This is in accordance with the fact that the three giant gravitons wrapping three independent $S^3$s in $S^5$ generically intersect in a point, and the degrees of freedom living at the intersection are $(0+1)$-dimensional.  Similarly, in the two-charge case, one expects a $(1+1)$-dimensional theory living at the intersection of two giant gravitons, and this nicely matches the appearance of a warped $\ads{3}$ metric in the decoupling limit.

Finally, we will discuss remaining open problems and possible interpretations of the results, and in the appendices collect some technical results and display a novel solution of type IIB supergravity which is tantalizingly close but not exactly identical to the metrics obtained in the decoupling limit of the two-charge solution, and which contains a warped $\ads{3}$ factor.

\section{Entropy of near-extremal $\ads{5}$ black holes}

Type IIB string theory in $\ads{5}\times S^5$ has a spectrum of charged black holes with a metric~\cite{superstar,cveticetal}:
\begin{eqnarray}
ds^2 &=& -\frac{\sqrt{\gamma}}{H_1H_2H_3}\,f\,dt^2 + \frac{\sqrt{\gamma}}{f}\,dr^2 + \sqrt{\gamma}\,r^2\,ds^2_{S^3} \nonumber \\
&& + \frac{1}{\sqrt{\gamma}}\sum_{i=1}^3 H_i\left(L^2\,d\mu_i^2 + \mu_i^2\left[L\,d\phi_i + (H_i^{-1}-1)\,dt\right]^2\right) \,,
\Label{eq:nonextremal2}
\end{eqnarray}
where
\be
H_i=1+q_i/r^2 \,, \quad
f=1-\mu/r^2+r^2\,H_1H_2H_3/L^2 \,, \quad
\gamma=H_1H_2H_3\sum_{i=1}^3 \mu_i^2/H_i \,,
\ee
and
\be
\mu_1 = \cos\theta_1~, \ \mu_2 = \sin\theta_1\,\cos\theta_2~, \ \mu_3 = \sin\theta_1\,\sin\theta_2 \,.
\Label{mui}
\ee
Here, $L = (4\pi g_s N)^{1/4} \ell_s$ is the AdS radius and $g_s N$ appears as the 't Hooft coupling $\lambda$ in the dual $SU(N)$ supersymmetric Yang--Mills gauge theory.

There is also a self-dual five-form field strength
$F^{(5)}=dB^{(4)}+\star dB^{(4)}$
with
\be \label{flux}
B^{(4)} = -\frac{r^4}{L} \gamma dt\wedge d^3\Omega - L\sum_{i=1}^3 q_i \mu_i^2 (Ld\phi_i - dt)\wedge d^3\Omega \,.
\ee

This metric (\ref{eq:nonextremal2}) is obtained by uplifting charged black hole solutions to five-dimensional gauged supergravity of the form
\begin{eqnarray}
ds_5^2 &=& -(H_1 H_2 H_3)^{-2/3} \, f \, dt^2 + (H_1 H_2 H_3)^{1/3} \, (f^{-1} dr^2 + r^2 d\Omega_3^2) \,, \nonumber \\
X_i &=& H_i^{-1} (H_1 H_2 H_3)^{1/3} \,, \quad A^i = {\tilde{q}_i \over r + q_i} dt \,, \quad (i = 1,\cdots, 4) \,,
\Label{5dsoln}
\end{eqnarray}
where $f$ and $H_i$ are defined as above, and
$X_i$ and $A^i$ are the scalars and gauge fields of five-dimensional gauged supergravity.
The five-dimensional Newton constant is
$G_5 = G_{10}/{\rm Vol}(S^5) = G_{10} / \pi^3 L^5 \simeq L^3/N^2$,
where we have used $G_{10} = 8\pi^6 g_s^2 \ell_s^8$.
The charge and mass of the black hole are
\begin{eqnarray}
\widetilde{q}_i &=& \sqrt{q_i(q_i+\mu)} \,, \Label{charge} \\
M &=& \frac{\pi}{4G_5} \left( \frac32 \mu + \sum_i q_i \right) \Label{mass} \,.
\end{eqnarray}
The parameters $q_i$ and $\mu$ have dimensions of length squared, so it convenient to define dimensionless quantities
\begin{equation}
q_i = L^2\,\hat{q}_i \,, \quad
\mu = L^2 \,\hat{\mu} \,; \quad
\hat{q}_i = {N_i \over N} \,,
\Label{dimensionless}
\end{equation}
where $N_i$ is the integral number of D-branes (giant gravitons) associated to each charge species \cite{superstar}.
The analysis of the microstates of the single R-charge extremal black holes has been carried out in the regime $\hat{q} = O(1)$ \cite{babel1}.
Most of the microstates cluster around a typical configuration of D-branes (giant gravitons).
Since one of our goals is to give a physical picture of the entropy of near-extremal black holes in terms of open strings on these D-branes, we will usually take $\hat{q}_i$ to be $O(1)$ --- {\em i.e.}, these parameters will not scale with $N$.
Finally, assuming the existence of a non-vanishing horizon $r_h$, the entropy of these black holes scales like:
\begin{equation}
S = {A \over 4 G_5} \sim \frac{N^2}{L^3}\,\sqrt{(r_h^2+q_1)\,(r_h^2+q_2)\,(r_h^2+q_3)}~,
\Label{eq:mchargeentropy}
\end{equation}
where $A$ is the area of the outer horizon and $r_h$ is the largest positive solution of $f(r) = 0$.

\paragraph{Single-charge black hole: }
The singly charged black hole is obtained by setting $H_2 = H_3 = 1$ in (\ref{5dsoln}).
For general values of $\mu$, the singularity at the origin is spacelike and becomes null in the extremal limit ($\mu = 0$), where it preserves sixteen supercharges and is one of the half-BPS solutions described in~\cite{llm};
following their conventions the Planck length is related to $\hbar$ in the dual gauge theory as $\ell_P^4 \leftrightarrow \hbar$.
The extremal black hole can be regarded as the backreacted geometry of a configuration of $N' = q\,N/L^2$ giant gravitons (D3-branes wrapping an $S^3$ in the $S^5$),
or alternatively as a configuration of $N$ dual giant gravitons (D3-branes wrapping an $S^3$ in the $\ads{5}$) \cite{superstar}.
The vast majority of bound states of giant gravitons having identical global charges source geometries that differ from each other at Planck distances.
Such fine distinctions among the geometries are lost in the coarse-graining that is implied by the semiclassical limit, which here corresponds to taking $N\to\infty$ while keeping $\hbar\,N \simeq L^4$ fixed~\cite{babel1,llm,quantgrav}.
Because the Ramond--Ramond potential for this solution to type IIB supergravity is independent of the non-extremality parameter $\mu$~\cite{superstar,cveticetal}, we can regard the non-extremal black holes as supersymmetry breaking deformations of the extremal half-BPS solutions.
In particular, departure from extremality does not alter the number of D-branes whose backreaction sources the black hole spacetime.
In the dual field theory, the black hole is described by a state of conformal dimension and R-charge
\begin{eqnarray}
&& \Delta = M \cdot L \simeq \frac{L^6}{G_{10}} \cdot \left( q + \frac{3}{2}\mu \right) \simeq N^2\,\hat{q}\left(1+\frac{3}{2}\frac{\hat{\mu}}{\hat{q}}\right) \,, \Label{eq:cd1} \\
&& J = \widetilde{q} \cdot {L \over G_5} \simeq N^2\,\hat{q}\,\sqrt{1+\hat{\mu}/\hat{q}} \approx N^2\,\hat{q}\,\left(1+ \frac{1}{2}\frac{\hat{\mu}}{\hat{q}}\right) \,. \Label{eq:rcharge1}
\end{eqnarray}
where the last expressions are valid in the near-extremal limit
\begin{equation}
\epsilon := \hat{\mu}/\hat{q} \ll 1 \,.
\label{eq:pt}
\end{equation}
The extremal $\Delta = J$ solutions are BPS.
Starting with a given BPS solution, variation of $\mu$ gives a one-parameter family of non-BPS black holes, in which the number of D-branes stays fixed at least near extremality.
Along this family of solutions, in the near-extremal limit ($\epsilon \ll 1$),
each additional unit of R-charge costs three units of conformal dimension:
\begin{equation}
\delta\Delta = \frac{3}{2}\,N^2\,\hat{\mu} \,, \quad
\delta J = \frac{1}{2}\,N^2\,\hat{\mu} \,.
\label{eq:def1}
\end{equation}
The outer radius of the black hole $r_h$ is given by the maximum real $r$ such that the function $f(r)$ in the metric \eref{5dsoln} vanishes:
\begin{equation}
4\,r_h^2 = -2(L^2 + q) + 2(L^2 + q)\,\sqrt{1 + \frac{4\,\mu\,L^2}{(L^2+q)^2}} \approx 4\,L^2\,\frac{\hat{\mu}}{1+\hat{q}} \,,
\Label{horizon}
\end{equation}
where the simplification applies in the near-extremal limit. 
When $\mu=0$, the radius of the horizon vanishes.
The gravitational entropy is given by
\be
S_{BH} \simeq \frac{r_h^2(r_h^2+q)^{1/2}}{G_5}
\approx N^2\,\hat\mu\,\frac{\sqrt{\hat{q}}}{1+\hat{q}} \,.
\Label{eq:sbh1}
\ee

\paragraph{Two-charge black holes: }
The two-charge black is obtained by setting one of the harmonic functions, say $H_1$, to unity in (\ref{5dsoln}).
The horizon occurs at
\begin{equation}
r_h^2 = \frac{L^2+q_2+q_3}{2}\,\left(-1 + \sqrt{1 + 4\,\frac{\mu\,L^2-q_2\cdot q_3}{(L^2+q_2+q_3)^2}}\right) \,.
\end{equation}
There is a critical value of the non-extremality parameter
\begin{equation}
\mu_{\text{crit}}= {q_2\,q_3 \over L^2}
\end{equation}
below which the horizon ceases to exists.
For $\mu > \mu_{{\rm crit}}$ the singularity is spacelike, for $\mu = \mu_{{\rm crit}}$ it is null, and when $0< \mu < \mu_{{\rm crit}}$ it is timelike and naked.
As $\mu$ approaches the critical point from above the horizon radius smoothly goes to zero.
At the {\it critical point}, the conserved charges are given by:
\begin{eqnarray}
\Delta_{\text{crit}} &=& M \cdot L = N^2\left(\hat{q}_2+\hat{q}_3 + \frac{3}{2}\,\hat{q}_2\hat{q}_3\right) \,, \\
J_2 &=& \tilde{q}_2 \cdot {L \over G_5} = N^2\,\hat{q}_2\,\sqrt{1+\hat{q}_3} \,, \\
J_3 &=& \tilde{q}_3 \cdot {L \over G_5} = N^2\,\hat{q}_3\,\sqrt{1+\hat{q}_2} \,.
\end{eqnarray}
Thus at the critical point $\Delta \neq J_2 + J_3$ and hence the solution is {\it not} BPS.
In the range $0 < \mu < \mu_{{\rm crit}}$ we have line of non-BPS naked singularities, while $\mu = 0$ describes a supersymmetric solution.
In section~4 we will propose a string theoretic resolution of these singularities.

One can define a {\it near-critical} limit as:
\begin{equation}
\mu = \mu_{\text{crit}} + \delta\mu \,, \quad 4\frac{\delta\mu\,L^2}{(L^2+q_2+q_3)^2}\ll 1 \,.
\label{eq:ncritical2}
\end{equation}
For generic charges $\hat{q}_2 \sim \hat{q}_3 \sim O(1)$, in this limit the horizon radius and the entropy scale as:
\begin{eqnarray}
r_h^2 &\sim& \frac{\delta\mu}{1+\hat{q}_2+\hat{q}_3} \,, \nonumber \\
S &\sim& N^2\,\left(\frac{\delta\hat{\mu}\,\,\hat{\mu}_{\text{crit}}}{1+\hat{q}_2+\hat{q}_3}\right)^{1/2}
= \ N^2\,\left(\frac{\delta\hat{\mu}\,\,\hat{q}_2 \, \hat{q}_3}{1+\hat{q}_2+\hat{q}_3}\right)^{1/2}
\approx N^2 (\delta\hat{\mu} \, \hat{q}_2 \hat{q}_3)^{1/2} \,.
\Label{2chargeentropy}
\end{eqnarray}
The square root scaling here with $\delta \mu$ differs from the linear scaling in the near-extremal limit for the single-charge black hole, and the last equality applies when the number of D-branes is much less than $N$ ({\em i.e.}, $N_i \ll N$).

\paragraph{Three-charge black holes: }
The three-charge black hole also has a critical value of the non-extremality parameter below which the solution displays a naked timelike singularity.
However, the situation is quite different from the two-charge case.
The equation for the horizon radius reads
\be \label{hor3}
r_h^4 - \mu r_h^2 +
\frac{1}{L^2}(r_h^2+q_1)(r_h^2+q_2)(r_h^2+q_3)=0 \,,
\ee
and there is a critical value $\mu_{\rm crit}$ below which this equation has no solution.
However, as $\mu\rightarrow \mu_{\rm crit}$ the horizon area remains of finite size ---
the entropy cannot be made arbitrarily small.
The explicit expression for $\mu_{\rm crit}$ is very cumbersome and can be obtained from the discriminant of (\ref{hor3}).
In the limit of small charges, the critical value of $\mu$ behaves as
\be
(\mu_{\rm crit} - \mu_c)^2
 \sim \frac{4 q_1 q_2 q_3}{L^2} \,,
\ee
where we have assumed that all charges are made small at the same rate
and $\mu_c = (q_1 q_2 + q_1 q_3 + q_2 q_3)/L^2$.
If we write $\mu=\mu_{\rm crit}+\delta \mu$ with $\delta\mu \ll \mu_{\rm crit}$, and consider a small charge limit,\footnote{
To be precise, we consider the limit $q_i \rightarrow \epsilon q_i$ and $\mu\rightarrow \epsilon^{3/2} \mu$, and send $\epsilon\rightarrow 0$.}
the entropy behaves as
\be \label{3ent}
S \sim N^2 \sqrt{\hat{q}_1 \hat{q}_2 \hat{q}_3} + \frac{N^2}{2} (\hat{q}_1\hat{q}_2 + \hat{q}_2 \hat{q}_3 + \hat{q}_3 \hat{q}_1) \left( 1 + \frac{\delta \hat{\mu}}{\sqrt{\hat{q}_1 \hat{q}_2 \hat{q}_3}} \right) +
O\left(\hat{q}^{5/2},\frac{\delta\hat{\mu}}{\hat{\mu}_{\rm crit}}\right) \,.
\ee
Surprisingly, in this limit, the leading term in the entropy is completely independent of $\delta \mu$.
At $\mu=\mu_{\rm crit}$ the horizon radius equals (again for small charges) $r_h^2=\mu_{\rm crit}/2$ with entropy $S \sim N^2 \sqrt{\hat{q}_1 \hat{q}_2 \hat{q}_3}$.
At this point the horizon radius is smaller than the AdS curvature radius, and one might expect that the behavior of the entropy as a function of $\mu$ will be modified in string theory, though our analysis of the decoupling limit of the three-charge case in section~\ref{threecharge} does not yet shed much light on this issue.

\subsection{Scales in the single-charge black hole}

The semiclassical analysis of black hole entropy that is described above is valid when stringy and quantum gravitational corrections are negligible because the horizon is sufficiently large.
This is the case when the length and curvature scales of the horizon exceed the string length ($\ell_s$) and the Planck length ($\ell_P \sim g_s^{1/4} \ell_s$).

The ten-dimensional horizon area of the single-charge black hole is $A_H \sim r_h^2 \sqrt{r_h^2 + q}\, L^5$.
Comparing to $\ell_s$ and $\ell_P$ gives:
\begin{eqnarray}
A_H &\gg& \ell_P^8: \quad \hat{\mu} \gg {1 \over N^2} \,, \\
A_H &\gg& \ell_s^8: \quad \hat{\mu} \gg {1 \over (g_s N)^2} \,.
\end{eqnarray}
This means that the black hole has a finite area in Planck units when the mass above extremality goes as $\delta M \sim L^2 \hat{\mu} / G_5$ (see (\ref{mass})).
Translating into dual CFT units, the horizon has finite area in Planck units if the deformation of the extremal state increases the conformal dimension (\ref{eq:cd1}) by
$\delta \Delta \sim \delta M \cdot L \sim N^2 \cdot \hat{\mu} \sim O(1)$,
while for a horizon that has finite area in string units we need
$\delta \Delta \sim {N^2\over (g_s N)^2} \sim {N^2 \over \lambda^2}$,
where $\lambda$ is the 't Hooft coupling constant of the dual Yang--Mills theory which is large but finite for the AdS/CFT duality to apply.
Next, consider the ten-dimensional curvature invariants ${\cal R} = R_{\mu\nu} R^{\mu\nu}$ and $K = R_{\mu\nu\rho\sigma} R^{\mu\nu\rho\sigma}$ evaluated at the black hole horizon.
To leading order in $\hat{\mu}$ at $\theta_1 > 0$:\footnote{
These expressions look singular as $\theta_1 \to 0$, but this is simply an order of limits issue.
Taking $\theta_1 \to 0 $ and then extracting the leading terms in $\hat\mu$ leads to a non-singular expression for the curvatures at any $\hat\mu > 0$.
Also the Ricci scalar vanishes smoothly in the extremal $\mu \to 0$ limit.}
\be
{\cal R} = \frac{2(1+\hat{q})(5-2\hat{q}+5\hat{q}^2)}{L^4\hat\mu\hat{q}\sin^2\theta_1} \,, \qquad
K = \frac{8(1+\hat{q})(4+5\hat{q}+4\hat{q}^2)}{L^4\hat\mu\hat{q}\sin^2\theta_1} \,.
\Label{curv}
\ee
These exceed the Planck and string scales when:
\begin{eqnarray}
K, {\cal R} &\gg& \ell_P^4: \quad \hat{\mu} \gg {1 \over N} \,, \\
K, {\cal R} &\gg& \ell_s^4: \quad \hat{\mu} \gg {1 \over (g_s N)} \,.
\end{eqnarray}
Thus to have horizon curvatures that are small in Planck units we need CFT deformations with $\delta\Delta \sim O(N)$, while curvatures that are small in string units require $\delta \Delta \sim {N^2 \over \lambda}$.

Putting these bounds together, to have a near-extremal black hole with a horizon that does not acquire quantum gravity and string corrections, we require that
\begin{equation}
{1 \over g_s N} \ll \hat\mu \ll \hat{q} \,.
\end{equation}
In the limit where the AdS/CFT correspondence applies, we take the large-$N$ limit with $g_s N$ held fixed and large.
Hence the lower bound can be regarded as a small number that is $O(1)$ in terms of scaling with $N$.
Finally, a near-extremal black hole with a horizon that is finite with respect to the AdS scale satisfies
\begin{equation}
{r_h \over L} \sim O(1) ~~~\Longrightarrow~~~ \hat{\mu} \sim O(1) \ll \hat{q} \,,
\end{equation}
where we used the fact that $\hat{q}$ does not scale with $N$ and the second inequality is simply the requirement that the black hole be near extremality.
In the limit where the AdS/CFT correspondence applies (large-$N$ with $L$ fixed but large), such black holes are reliably described in five-dimensional supergravity, so long as instabilities to Hawking radiation and localization on the $S^5$ \cite{BDHM} do not set in.
These instabilities appear when the black hole horizon is smaller than the AdS scale \cite{cveticgubser,cejm,gubsermitra,gubsermitra1,yaya} by a factor that is $O(1)$ (and not parametrically smaller in $N$ or $g_s$).
Thus the near-extremal gravitational entropy calculated above can only match the degeneracy in the dual field theory in some range:
\begin{equation}
{1 \over g_s N} \ll \delta \ll \hat{\mu} \ll \hat{q} \sim O(1) \,,
\end{equation}
where $\delta$ is an $O(1)$ number that could be determined by a detailed stability analysis in gravity.

\section{Counting states}
\label{sec:fieldtheory}

We will provide evidence that the microstates of the near-extremal black hole can be thought of as a dilute gas of defects distributed within a heavy BPS operator which in turn creates one of the typical microstates of the extremal black hole.
This picture is motivated by the observation described in the previous section that the underlying distribution of D-branes that source the spacetime is not modified in the near-extremal black hole, suggesting that the entropy of the near-extremal black hole arises from a gas of open strings attached to these branes.
Techniques for enumerating such open string states and studying their dynamics have been developed in \cite{BHNL,davidemergence,Berensteinetal,BBFH,MelloKoch}.
We will first provide a back-of-the-envelope combinatorial estimate of the number of ways of distributing defects on a large BPS operator, and then we will derive the same functional form for the degeneracy from the exact $\CN=4$ super-Yang--Mills partition function at zero coupling.
In this section we will only study the single-charge black hole, and it should be kept in mind that in order to match with the black hole an extrapolation to finite coupling is necessary;
in the absence of supersymmetry the degeneracy can change with the coupling as the operators are not protected from obtaining anomalous dimensions, but it is interesting to see how far we can go by working with the weakly coupled description.

\subsection{The typical states of extremal black holes}

The extremal black hole ($\mu=0$) preserves sixteen of thirty-two supercharges.
Such half-BPS configurations are amenable to an exact analysis in both field theory and bulk supergravity using the techniques of~\cite{BBNS,cjr,ber,llm}.
In the $\CN=4$ super-Yang--Mills gauge theory, half-BPS multiplets transform in the $[0,p,0]$ representation of the $SO(6)$ R-symmetry.
The highest weight state in each half-BPS multiplet is a multi-trace operator in a single complex chiral superfield $Z$.
Because $Z$ carries a single unit of $U(1)$ R-charge and also one unit of conformal dimension, a polynomial in $Z$ automatically satisfies the half-BPS condition that $\Delta=J$.
The structure of the half-BPS operators can be completely studied in a reduction of the Yang--Mills theory to a Hermitian matrix model with a harmonic oscillator potential \cite{cjr, ber}.
The eigenvalues of this system are governed by the dynamics of $N$ fermions in a harmonic potential.
The ground state of the $N$ fermi system encodes empty $\ads{5}\times S^5$, while excited states correspond to half-BPS geometries via an explicit dictionary~\cite{llm,babel1, davidemergence, othershalfBPS}.

The data about the energy levels of the fermions are conveniently encapsulated in a Young diagram with $N$ rows.
The length $r_k$ of the $k$-th row denotes the excitation energy $E_k$ of the $k$-th fermion above its ground state energy $E_k^g$ (in conventions in which the frequency of the harmonic oscillator satisfies $\omega=1$):
\be
r_k = \frac{1}{\hbar} (E_k - E_k^g) \,.
\ee
The number of boxes in the Young diagram determines the conformal dimension $\Delta$ of this half-BPS operator.
The different Young diagrams are orthogonal configurations.

The Young diagram has a nice interpretation in terms of the D-branes that source the bulk geometry.
If we read the diagram from left to right, the columns represent giant gravitons, D3-branes that wrap an $S^3 \subset S^5$.
The maximum angular momentum of any of these giant gravitons is $N$, but their number is unconstrained.
If we read the diagram from top to bottom, the rows represent dual giant gravitons, D3-branes that wrap an $S^3 \subset \ads{5}$.
We can have at most $N$ of these because each of the dual giant gravitons is stabilized by one unit of flux.
The angular momentum of the dual giant gravitons (row length) is unbounded.

The extremal black hole corresponds to a geometry with a fixed number $N'$ of giant gravitons, where $N'$ is $O(N)$.
This means that the conformal dimension $\Delta$ of the dual operator lies between $N'$ and $N\cdot N'$.
We can consider the general shape of a Young diagram with a number of boxes $\Delta$ that lies within this interval.
That is to say, we enumerate partitions of $\Delta$ into at most $N$ parts such that the largest part is $N'$.
Almost all partitions lie along a limit shape that maximizes the entropy of the partition~\cite{babel1}.
The number of partitions that deviate significantly from this limit shape is exponentially small.
For $\Delta=\frac12 NN'$ the mean field curve describing the typical Young diagram is a triangle with sides of length $N$ and $N'$, and the entropy is computed to be
\be
S = \log\left( \frac{(\lambda + \sigma)^{\lambda + \sigma}}{\lambda^\lambda \sigma^\sigma} \right) \sqrt\Delta \,, \quad
\quad
\Delta = \frac12\, N\, N' \,,
\quad
\lambda = \sqrt{\frac{2\,N'}{N}} \,,
\quad
\sigma = \sqrt{\frac{2\,N}{N'}} \,.
\ee

The limit shape configuration determines the metric for the bulk geometry.
The slope of the curve $N'/N$ determines the distribution of the fermions in their phase space.
This information recovers the metric in the bulk.
Because the regular boundary conditions are not satisfied, the spacetime geometry in the semiclassical limit is singular.
In fact, it is the metric \eref{eq:nonextremal2} with $q_2=q_3=\mu=0$~\cite{babel1}.
Equating the microcanonical entropy $S$ to the bulk entropy $S_{BH} \simeq r_h^3/G_5$, the horizon size of the extremal black hole scales as a negative power of $N$ and is therefore not detectable semiclassically.

\subsection{The non-extremal black hole as a gas of defects}

In analyzing microstates of the non-extremal black hole, which we treat as perturbations of the extremal operators, it is convenient to work in a basis where operators are written as products of traces.
To map from the free fermion picture to the multi-trace picture, a Schur polynomial is associated to each Young diagram as follows:
\be
\sum_{\{a_i\}} C_{\{a_i\}} \prod_i \tr{Z^{a_i}} \,,
\Label{tracebasis}
\ee
where $Z$ is a complex scalar field of the Yang--Mills theory, the $\{a_i\}$ are partitions of $\Delta$ subject to the constraint that $a_i \le N$ to satisfy $U(N)$ trace relations, and the $C_{\{a_i\}}$ are certain coefficients constructed from characters of the symmetric group \cite{cjr,ber}.
Thus, the typical half-BPS state that is dual to an extremal black hole microstate can be expressed in terms of a particular sum of multi-trace operators following (\ref{tracebasis}).

Given such a typical state in the trace basis, we can introduce a diffuse gas of randomly distributed supersymmetry breaking defects.
The number of inequivalent ways of introducing these defects should account for the entropy of the non-extremal black hole.
From the spacetime perspective, these defects correspond to open strings attached to the D-branes creating the black hole \cite{BHNL,davidemergence,Berensteinetal,BBFH,MelloKoch}.

The defects that could be introduced into a half-BPS operator can be constructed from products of the elementary defects such as the ones listed below with their quantum numbers:
\vspace*{.5cm}
\newline
\begin{tabular}{c|cccccccc|}
              & $M_{ij}$ & $\lambda_{\alpha i}$  & $F_{(\alpha\beta)}$ & $(\nabla\nabla)_{(\alpha\beta)}$ & $\nabla^2$ & $(\nabla\bar\lambda)_\alpha^i$ & $(\bar \lambda \bar \lambda)^{ij}$ & $(FF)$ \\
    \hline
     $\Delta$ & $1$ & $3/2$ & $2$ & $2$              & $2$        & $5/2$          & $3$                 & $4$    \\
     $j$      & $0$ & $1/2$ & $1$ & $1$              & $0$        & $1/2$          & $0$                 & $0$    \\
     $[k,p,q]$& $[0,1,0]$ & $[0,0,1]$ & $[0,0,0]$ & $[0,0,0]$          & $[0,0,0]$        & $[1,0,0]$          & $[0,1,0]$                 & $[0,0,0]$    \\
     \hline
\end{tabular}
\vspace{.5cm}
\newline
$M_{ij}$ are the adjoint scalars and $i,j$ are $SU(4) \sim SO(6)$ indices, $\lambda_{\alpha i}$ are gauginos, $F_{\alpha\beta}$ is the $SU(N)$ field strength, and $\nabla$ is a gauge-covariant derivative.
Also $j$ is the rotational quantum number on the sphere where the Yang--Mills theory is defined and $[k,p,q]$ indexes the representation of the defect under the $SO(6)$ R-symmetry group.
Now, according to (\ref{eq:def1}), turning on $\hat{\mu}$ while keeping $q$ fixed means adding $\delta \Delta$ and $\delta J$ in the ratio $3:1$.
It is interesting that there are defects such as $(\bar \lambda \bar \lambda)^{ij}$ that directly achieve this goal, since this might suggest that this defect will be the most ``efficient'' way of achieving non-extremality and will dominate the degeneracy.
However, there is no obvious {\em a priori} reason why increasing $\mu$ at fixed $q$ should be a distinguished one-parameter family of solutions in the total space parameterized by both $\mu,q$.
Therefore, one should really count all possible distributions of arbitrary impurities over half-BPS states.
We will see later in this section that the final answer, {\em i.e.}, the degeneracy $S(\Delta,J)$, does not depend very much on whether one includes just a few impurities or a large number of them, or on whether one only considers impurities with $\delta\Delta:\delta J=3:1$.
We will therefore focus on a direct counting of $S(\Delta,J)$ in the remainder of this section.

\subsection{Combinatorial analysis}

In this section we will directly estimate the number of operators of given $\Delta,J$ with $\Delta\sim N^2$.
Consider a half-BPS operator of dimension $O(N^2)$.
Typically it will consist of a sum of parts, each of which is a product of many single-trace components.
Let us take a single one of these,\footnote{
The fact that there is an underlying ensemble of $O(e^{\sqrt\Delta})$ half-BPS operators of conformal dimension $\Delta$ is a subleading contribution to the entropy compared to the distribution of the defects in a typical state.}
which is a product of single-trace pieces,
\be
{\cal O} = {\rm tr}(Z^{a_1}) {\rm tr} (Z^{a_2}) \ldots {\rm tr}(Z^{a_M}) \,.
\Label{multitraceop}
\ee
We want to distribute a certain number of impurities over this operator, breaking supersymmetry.
However, instead of distributing arbitrary impurities we will for simplicity only distribute $s$ types of operators with $\Delta=1$ and $J=0$ over ${\cal O}$.
While this is a drastic simplification, one might expect it to yield the correct overall scaling.
Adding $\epsilon \Delta$ such defects into the traces appearing in the half-BPS operator would increase the conformal dimension of the operator by an amount $\epsilon \Delta$.
To remain in the near-extremal black hole limit we will take $\epsilon \ll 1$ and keep $\hbar\,N$ fixed while sending $N\to \infty$.

If we distribute all these impurity operators over ${\cal O}$, we will be adding a certain number of them to each single-trace component.\footnote{
We ignore ``pure'' defects, which is to say we incorporate defects into the single-trace components of ${\cal O}$, but do not account for new single-trace components (gauge invariant operators) built exclusively from the defects.
This undercounting is parametrically small in the large-$N$ limit.}
Suppose we add $n_{i,j}$ operators of type $j$ to ${\rm tr}(Z^{a_i})$. So $j=1,\ldots,s$ and $i=1,\ldots,M$.
Focusing on a single-trace component, \Polya\ counting tells us precisely how many ways there are of doing this.
However, for our purposes a simpler counting is sufficient.
Simple combinatorics tell us that the number of ways to distribute defects is approximately
\be \label{eq01}
\prod_i \frac{1}{a_i+\sum_j n_{i,j}} \left( \begin{array}{cc} a_i + \sum_j n_{i,j} \\ n_{i,1} \,\,\, n_{i,2} \,\,\, \cdots \,\,\, n_{i,s} \end{array} \right) \,,
\ee
Here, the factor of $\frac{1}{a_i+\sum_j n_{i,j}}$ appears due to the cyclicity of each of the traces.
The expression (\ref{eq01}) undercounts insertions of impurities that have additional symmetries for which not all cyclic shifts yield new configurations, but that is a subleading contribution to the overall counting.
Incidentally, this expression (\ref{eq01}) is exactly the first term in the usual \Polya\ expression.
If we denote
\be
m_i = \sum_j n_{i,j} \,,
\ee
then (\ref{eq01}) can be rewritten exactly as
\be \label{eq02}
\prod_i \frac{s^{m_i}}{a_i+m_i} \left( \begin{array}{cc} a_i + m_i \\ m_i \end{array} \right) \,.
\ee
We want to sum this and keep $\sum m_i = \epsilon \Delta$ fixed.
We can equivalently also sum over all $m_i$ and at the end simply pull out the relevant power of $s$.
The sum over all $m_i$ can be done because of the identity
\be \label{pp}
\sum_{m=0}^{\infty} \frac{s^m}{a+m} \left( \begin{array}{cc} a + m
\\ m \end{array} \right)  = \frac{1}{a(1-s)^a} \,,
\ee
and using this and upon extracting the appropriate power of $s$, we find
\be
\label{eq03} \frac{s^{\epsilon \Delta}}{\prod_i a_i} \left( \begin{array}{cc} \sum_i a_i + \epsilon \Delta -1 \\ \sum_i a_i-1 \end{array} \right) \,.
\ee
Since $\sum_i a_i =\Delta$, we can approximate the binomial coefficient and the result is the degeneracy
\be \label{eq04}
\frac{s^{\epsilon \Delta}}{\prod_i a_i} \frac{{\Delta}^{\epsilon \Delta}}{(\epsilon \Delta)!} \,.
\ee
The factor $(\prod a_i)^{-1}$ needs to be averaged over all possible states, but this will never produce a factor that depends on $\epsilon$.
Also, as this leading order result is independent of the precise distribution of trace lengths in the operator, we need not be concerned that we did not impose the $U(N)$ trace relations and specify the $a_i$ in (\ref{multitraceop}) to be exactly the distribution of traces appearing in the typical black hole microstate.
Using Stirling's approximation of the factorial, we obtain
\be
S \sim -\Delta \epsilon \log \epsilon + \epsilon \Delta (\log s+1) + \ldots \,.
\Label{eq05}
\ee
To match this with the near-extremal black hole we may use (\ref{eq:cd1}) and (\ref{eq:rcharge1}) to write:
\begin{eqnarray}
J_{{\rm tot}} &=& N^2(\hat{q} + {1 \over 2} \hat{\mu} ) = \Delta \,,
\Label{jtot}  \\
\Delta_{{\rm tot}} &=&N^2( \hat{q} + {3 \over 2} \hat{\mu})  = \Delta + \epsilon \Delta \,.
\label{deltatot}
\end{eqnarray}
Using this, the estimate (\ref{eq05}) can be written as
\begin{equation}
S \sim N^2 \hat{\mu} \log(\hat{q}/\hat{\mu}) + N^2\hat{\mu} \log(s+1) \,.
\end{equation}
While this matches the $N$ dependence of the black hole entropy (\ref{eq:sbh1}), the functional dependence on $\hat{\mu}$ and $\hat{q}$ is not the same.
In going from weak to strong coupling the degeneracy appears to finitely renormalize in a way that depends on the distance from extremality.
Below we will verify that the above estimate is accurate via a partition function calculation.

\subsection{Eigenvalue density analysis}

The exact zero coupling partition function of ${\cal N}=4$ Yang--Mills theory on $S^3$ in the presence of chemical potentials for the $U(1)$ charges is:
\bea
Z & = & {\rm Tr}(x^{\Delta} q_1^{J_1} q_2^{J_2} q_3^{J_3} )
\label{genfie}
\\ & = & \int [dU] \exp \left\{ \sum_{m=1}^{\infty} \frac{1}{m} (z_B(x^m,q_1^m,q_2^m,q_3^m) + (-1)^{m+1} z_F(x^m,q_1^m,q_2^m,q_3^m)) {\rm tr}(U^m) {\rm tr}(U^{\dagger})^m \right\} \,, \nonumber
\eea
where $\int [dU]$ is an integral over $U(N)$ with the boson and fermion one-particle partition functions $\{z_B,\,z_F\}$ given by
\bea
z_B(x,q_1,q_2,q_3) & = & \frac{6x^2-2x^3}{(1-x)^3} + \frac{x+x^2}{(1-x)^3}(q_1 + q_1^{-1} + q_2 + q_2^{-1} + q_3 + q_3^{-1}) \,, \\
z_F(x,q_1,q_2,q_3) & = & \frac{2x^{3/2}}{(1-x)^3}(q_1^{1/2} + q_1^{-1/2})(q_2^{1/2}+q_2^{-1/2}) (q_3^{1/2} + q_3^{-1/2}) \,.
\eea
This expression follows directly from the results in
\cite{harvard}; more detailed discussions can, for example, be
found in \cite{yaya,harmark}.

If, instead of working in terms of the $U(N)$ matrices, we use their eigenvalues, the partition function becomes:
\be
\label{eq41} Z= Z_0 \int \prod \frac{d\theta_i}{2\pi} \exp \left( -\sum_{i\neq j} V(\theta_i-\theta_j) \right) \,,
\ee
where
\be
Z_0 = \frac{1}{N!}\exp \left\{ \sum_{m=1}^{\infty} \frac{N}{m} (z_B(x^m,q_1^m,q_2^m,q_3^m) + (-1)^{m+1} z_F(x^m,q_1^m,q_2^m,q_3^m)) \right\} \,,
\ee
and
\be
V(\theta)=\log 2 + \sum_{m=1}^{\infty} \frac{1}{m} (1-z_B(x^m,q_1^m,q_2^m,q_3^m)-(-1)^{m+1} z_F(x^m,q_1^m,q_2^m,q_3^m)) \cos (m\theta) \,.
\ee
Following \cite{harvard}, in the large-$N$ limit, we can replace the integral over eigenvalues by an integral over an eigenvalue density $\rho(\theta)$, which we assume to be normalized $\int d\theta\ \rho(\theta)=1$.
If we also denote $\rho_m=\int d\theta\ \rho(\theta) \cos m\theta$ and $V_m=\int d\theta\ V(\theta) \cos m\theta$, then the partition function becomes
\be \label{genfie2}
Z=Z'_0 \int \prod_{m} d\rho_m \exp\left( -\frac{N^2}{2\pi} \sum_{m=1}^{\infty} |\rho_m|^2 V_m \right) \,,
\ee
with
\be
V_m={2\pi \over m} (1-z_B(x^m,q_1^m,q_2^m,q_3^m)-(-1)^{m+1} z_F(x^m,q_1^m,q_2^m,q_3^m))
\ee
and $Z'_0$ a constant independent of $x,q_i$ which we will ignore from now on.
In the remainder of this section we will put $q_2=q_3=1$, as we are interested in counting states as a function of $\Delta$ and just a single $U(1)$ charge $J$, so that
\begin{equation}
Z = {\rm Tr}(x^{\Delta} q^{J}) \,.
\end{equation}
Notice that we can rewrite the partition function as
\be
Z= {\rm Tr}( (xq)^{(\Delta+J)/2} (xq^{-1})^{(\Delta-J)/2} ) \equiv {\rm Tr}( q_+^{\Delta+J} q_-^{\Delta-J} ) \,,
\ee
where we have defined
\be
q_+=\sqrt{xq} \,,\qquad q_-=\sqrt{x/q} \,.
\ee
From this expression and the BPS condition $\Delta\geq |J|$, it is clear that the partition function only exists for $q_{\pm}\leq 1$, in other words for $x\leq q \leq x^{-1}$.
The BPS partition function is recovered in the limit
\begin{equation}
  q_- \to 0 \,, \quad q_+ \,\,\, \text{fixed} \,,
\end{equation}
where it becomes
\begin{equation}
  Z_{BPS} = {\rm Tr}_{\Delta=J}q_+^{2J} \,.
\end{equation}
The fixed parameter $-1/\log q_+$ corresponds to the ``fictitious'' temperature used in \cite{babel1}.
Furthermore, the superstar geometry was reproduced in the semiclassical limit as the geometrical realization of typical states in an ensemble with $q_+\to 1$ (infinite ``fictitious'' temperature).
In appendix~\ref{halfbps}, we explicitly demonstrate how one recovers the exact counting of half-BPS states from (\ref{genfie}).

Once we move away from extremality, we need to study $Z(x,q)$ as a function of both $x$ and $q$, as the entropy $S(\Delta,J)$ will be obtained as the Legendre transform of $\log Z$.
Instead of working with $x$ and $q$, it will be more convenient to work with $q_{\pm}$.
As was shown in \cite{harvard}, $\CN=4$ Yang--Mills theory has a phase transition in the large-$N$ limit even at zero coupling.
This phase transition separates a confining phase with the free energy independent of $N$ from a deconfined phase with the free energy proportional to $N^2$.
(For a discussion of this phase transition in the presence of chemical potentials, see \cite{yaya}.)
For our discussion it will be important to know in which phase we are.
Since we are interested in a regime where $\Delta\sim N^2$, it turns out that we are very close to the transition line, but as we will demonstrate the entropies below and above the transition line have the same qualitative behavior.
This is reminiscent of the Horowitz--Polchinski correspondence principle \cite{hopo} and is quite possibly the weakly coupled version of it.

The phase transition takes place along the line
\be
z_B(q_+,q_-)+z_F(q_+,q_-)=1 \,.
\ee
This line is contained entirely inside the region $q_{\pm}\leq 1$ in which the partition function exists, and in this region the one-particle functions $z_B(q_+,q_-)$ and $z_F(q_+,q_-)$ are strictly monotonic functions of both $q_+$ and $q_-$.
Below the phase transition, in the confining phase, all $V_m>0$ and the partition function is dominated by a saddle point which corresponds to the constant eigenvalue density.
On the other hand, above but close to the phase transition line, the partition function is dominated by an eigenvalue density $\rho(\theta)=\frac{1}{2\pi}(1+\cos\theta)$.

We first look at the case where we are in the deconfined phase.
Very near to the phase boundary one finds (similarly to \cite{harvard}) that the partition function behaves as
\be \label{deconf}
\log Z \sim \frac{N^2}{4} (z_B(q_+,q_-)+z_F(q_+,q_-)-1) \,.
\ee
On the other hand, in the confining phase we find by integrating out $\rho_m$ that
\be \label{conf}
\log Z \sim -\sum_{m=1}^{\infty}
\log(1-z_B(q_+^m,q_-^m)-(-1)^{m+1} z_F(q_+^m,q_-^m) ) \,,
\ee
and very close to the phase boundary this is dominated by the term with $m=1$ which diverges there.

Now we consider a point $(q_+,q_-)$ very close to a point $(q_+^0,q_-^0)$ on the phase boundary.
We can write
\be \label{taylor}
z_B(q_+,q_-)+z_F(q_+,q_-) \sim 1 + a_0(q_+-q_+^0) + b_0(q_--q_-^0) + \ldots \,,
\ee
and from this we obtain the entropy as the Legendre transform
\be
S(\Delta,J)=\log Z - (\Delta+J)\log q_+ - (\Delta-J)\log q_-
\ee
of $\log Z$.
In the deconfining phase we get
\be
S\sim (\Delta+J) - \frac{N^2}{4}q_+^0  a_0 - (\Delta+J) \log\frac{4(\Delta+J)}{N^2 a_0} + (\Delta-J) - \frac{N^2}{4} q_-^0 b_0 - (\Delta-J)\log\frac{4 (\Delta-J)}{N^2 b_0} \,.
\ee
This result is only valid in the regime
\be \label{jan6}
\Delta+J \geq \frac{N^2}{4}q_+^0  a_0 , \qquad \Delta-J \geq \frac{N^2}{4} q_-^0 b_0 \,.
\ee
As we vary $(q_+^0,q_-^0)$, these bounds represent the phase boundary in the $(\Delta,J)$ plane which separates the confining from the deconfining phase.

In the confining phase we find, on the other hand, that the relation between $q_{\pm}$ and $\Delta,J$ reads
\be \label{jan7}
q_-=\frac{(a_0q_+^0 + b_0q_-^0)(\Delta-J)}{b_0(1+2\Delta)} \,,\qquad
q_+=\frac{(a_0q_+^0 + b_0q_-^0)(\Delta+J)}{a_0(1+2\Delta)} \,,
\ee
and that the entropy is given by
\be \label{ent3}
S \sim  (2\Delta+1)\log \left(\frac{2\Delta+1}{a_0q_+^0+b_0q_-^0} \right) - (\Delta+J)\log\frac{\Delta+J}{a_0} - (\Delta-J)\log\frac{(\Delta-J)}{b_0} \,.
\ee
If we insert the values for $\Delta,J$ from (\ref{jan6}) into (\ref{jan7}), we find that
\be
q_+=q_+^0-{O}(N^{-2}),\qquad q_-=q_-^0-{O}(N^{-2}) \,.
\ee
Therefore, we find that the results from the analysis of the confining phase are reliable up to a distance of order $N^{-2}$ away from the phase boundary.
There is a very small crossover phase of width $N^{-2}$, and above the phase boundary the results from the deconfining phase analysis are reliable.
The results for the entropy from the deconfining phase and confining phase match to leading order at the phase boundary
(both are equal to $-(\Delta+J) \log q_+^0 - (\Delta-J) \log q_-^0$),
which again is reminiscent of the Horowitz--Polchinski correspondence principle \cite{hopo}.

The above analysis of the confining phase will also break down once we approach the half-BPS point on the phase boundary where $q_-=0$ while $q_+=1$.
Here, (\ref{conf}) cannot be approximated by the $m=1$ term alone.
For $q_+$ very close to one and $q_-$ very close to zero $\log Z$ behaves in the confining phase as
\be \label{newlog}
\log Z \sim -\log\left( 2(1-q_+) - 16 q_-\right) + \frac{\pi^2}{12 (1-q_+)} \,.
\ee
The first term is the leading behavior of the $m=1$ term in (\ref{conf}) and a similar logarithmic divergence appears all along the phase boundary.
The second term is the leading behavior of the remaining terms $-\sum_{m=2}^{\infty} \log V_m$ in (\ref{conf}).
Recall that the entropy was obtained from the Legendre transform
$S(\Delta,J)=\log Z - (\Delta+J)\log q_+ - (\Delta-J)\log q_-$.
Solving for $q_{\pm}$ we find, up to subleading corrections, that they are given by
\bea
\frac{\pi^2}{12(1-q_+)^2} + \frac{\Delta-J}{1-q_+} - (\Delta+J) & = & 0 \,, \\
\frac{(1-q_+)}{8}\left( 1-\frac{1}{\Delta-J} \right) & = & q_- \,.
\eea
From these equations we see that there are two regimes, one where $(\Delta-J)^2 \ll (\Delta+J)$ and where $(1-q_+)\sim 1/\sqrt{\Delta+J}$, and one regime where $(\Delta-J)^2 \gg (\Delta+J)$ and $(1-q_+) \sim (\Delta-J)(\Delta+J)^{-1}$.
In the first regime, which is very close to the half-BPS limit, the entropy is given to leading order by
\be \label{conste}
S\sim \pi\sqrt{\frac{\Delta+J}{3}} \,,
\ee
whereas in the second regime it is given to leading order by
\be
S\sim (\Delta-J) \log\left(  \frac{\Delta+J}{\Delta-J} \right) \,.
\ee
In the first regime the leading contribution to the entropy is simply coming from the number of half-BPS states, and the additional contribution coming from the impurities can be neglected.
In the second regime this is no longer the case.
The crossover between the two regimes takes place when $(\Delta-J)^2$ is of order $(\Delta+J)$, and the crossover regime will therefore have size $\sim N^{-1}$.

\paragraph{Summary: }
We have studied the partition function and the entropy close to the confinement/deconfinement phase transition line, and we have found three regimes:
close to the phase transition line in the deconfined phase, close to the phase transition in the confined phase, and close to the half-BPS point.
These are all relevant in different regimes, and there is a smooth crossover behavior between them.
If $\Delta\pm J$ both scale as $N^2$, the entropy is linear in $\Delta\pm J$.
This result is even valid if $\Delta\pm J$ are proportional to large constants times $N^2$, but to see this one needs to study the high-temperature behavior which requires a separate analysis.
If $(\Delta+J) \ll (\Delta-J)^2 \ll (\Delta+J)^2$, then we are always in the confined phase but the results from the near-half BPS regime and the regime close to the phase boundary (\ref{ent3}) agree:
the entropy behaves as (assuming $\Delta+J\sim N^2$)
\be
S \sim (\Delta-J) \log \left({\Delta+J \over \Delta-J} \right)
\sim  N^2 \, \hat{\mu} \log (\hat{q} / \hat{\mu}) + N^2 \, \hat{\mu} \, \log 2 \,,
\Label{fine}
\ee
where we have used (\ref{jtot}) and (\ref{deltatot}) in writing the last equality.
This result is robust, as one can easily check that including higher terms in the Taylor expansion of $z_B+z_F$ will not affect the leading behavior of the entropy.
Finally, once $(\Delta-J)^2 \ll (\Delta+J)$, we are very close to the half-BPS point, and the entropy is dominated by that of half-BPS states as given in (\ref{conste}).

The result for the entropy (\ref{fine}) agrees with the leading behavior we obtained from the combinatorial analysis (see (\ref{eq05})), and we have therefore justified the interpretation of the microstates as open strings propagating on a background of half-BPS D-branes.
Following the discussion below (\ref{eq05}), we also see that the free field theory reproduces the $N$ scaling of the gravitational entropy in the regime $\hat{\mu}, \hat{q} \sim O(1)$ where the calculation applies, but not the functional dependence on the distance from extremality.

It is interesting that free field theory is sufficient to recover the scaling behavior in $N$ of the entropy of the non-extremal single-charge black hole in ${\rm AdS}_5$ given that the associated operators are not protected by supersymmetry.  This lends some credence to the proposal in \cite{babel1} that very heavy operators of conformal dimension $O(N^2)$ associated to black holes  might enjoy a kind of  ``almost-non-renormalization'' theorem.



\section{Decoupling Limits}

In the previous section we tried to account for the entropy of near-extremal black holes in two ways.
Firstly, we simply computed a partition function in the Yang--Mills theory.
Secondly, we counted defects inserted in the half-BPS state, which can be regarded as counting open strings on the D-branes sourcing the spacetime \cite{BHNL,davidemergence,Berensteinetal,BBFH,MelloKoch}.
These approaches agreed in their estimate of the entropy.
While we did these calculations for the single-charge superstar, the partition function calculation can be generalized in a fairly straightforward manner to the two- and three-charge cases with similar results.

However, the approach of counting defects, {\em i.e.}, open strings on giant gravitons, requires more thought in the two- and three-charge cases.
This is because the multi-charge black holes contain multiple species of giant gravitons that are oriented differently in the $S^5$ in $\ads{5} \times S^5$, and thus contain new kinds of open strings that are stretched between the different species of branes.
For the two-charge case, the two species of branes intersect on circles, and so if the strings stretched between these branes dominate, the entropy should be explained by an effective two-dimensional theory \cite{gubserheckman}, and be proportional to the number of intersections between the two brane species.
Indeed, the near-extremal entropy formula (\ref{2chargeentropy}) has a dependence on the product of the charges, $\hat{q}_2\, \hat{q}_3$, which is proportional to the intersection number.
Likewise for the three-charge case, the three kinds of branes mutually intersect at a point and if the degrees of freedom at this intersection dominate, the entropy should be explained by a $(0+1)$-dimensional theory, and be proportional to the product of all three charges.
Indeed, the latter is true for the near-extremal black hole: see (\ref{3ent}).
Finally, the square root form of (\ref{eq:mchargeentropy}) suggests that the entropy in all cases is associated to a partition of integers, possibly arising from an underlying conformal field theory.
Thus the entropy formul\ae\ point towards the existence of effective $(1+1)$- and $(0+1)$-dimensional conformal field theories that explain the entropy of two- and three-charge black holes in $\ads{5}$.\footnote{
See \cite{gubserheckman} for an attempt to directly analyze the theory on the intersection of a pair of giant gravitons to account for the superstar entropy.}
To test this, we can look for the $\ads{3}$ and $\ads{2}$ gravity duals of these effective field theories by taking decoupling limits that isolate the theories living on the intersections of giant gravitons.

Recall first that all the giant gravitons sourcing the black hole geometries are not located at the same point in the $S^5$ factor of $\ads{5} \times S^5$ \cite{superstar}.
To review, there are three kinds of giant gravitons, depending on the angular directions $\phi_i$ along which they move (the R-charge they carry).
In terms of the functions $\mu_i$ appearing in (\ref{eq:nonextremal2}), each giant of the i$^{\rm th}$ kind has a radius
\be
\rho_i = L\,\sqrt{1-\mu_i^2} \,,
\Label{giantsize}
\ee
and so its size depends on its location on the five-sphere.
The density of giants characterizing these distributions in each direction is \cite{superstar}
\begin{equation}
  \frac{dN_i}{d\rho_i} = 2N\,\frac{q_i}{L^4}\,\rho_i \,.
  \label{giantdist}
\end{equation}
Integrating this, we learn that the charge parameters $(q_i/L^2)$ controlling the black holes configurations are related to the integral number of D-branes by
\begin{equation}
  \frac{q_i}{L^2} = \frac{N_i}{N} \,.
\Label{totalgiants}
\end{equation}

We seek decoupling limits in which gravity in the near-brane geometry is dual to the low-energy field theory on the D-branes themselves.
Imitating the decoupling procedure that led to the AdS/CFT correspondence \cite{juanAdS}, we will take the string length $\ell_s$ to zero while keeping the energies of open strings relevant for the entropy fixed.
On the D-branes, this gives a theory of the lightest open string modes decoupled from gravity.
If $\Delta s$ is the length of an open string stretched between a pair of branes, its mass is $\Delta s / \ell_s^2$.
Hence, we must require
\begin{equation}
\ell_s \to \epsilon \, \ell_s \,; \qquad \Delta s \to \epsilon^2 \, \Delta s
\end{equation}
as $\epsilon \to 0$ in order to keep masses fixed.
Since the giant gravitons in the superstar solutions are at the origin of $\ads{5}$ and are distributed over the $S^5$, the decoupling limit in spacetime will have to focus in on $r^2 \sim 0$ and onto fixed angles on $S^5$ to achieve $\Delta s \sim \epsilon^2$.
Furthermore, since the giant gravitons are moving at the speed of light around the angular directions $\phi_i$ it will be generally necessary to also focus in on a null line to display the region where stretched open strings remain light.\footnote{
Giant gravitons are massive particles and strictly speaking their velocity is less than the speed of light.
However, they have an equation of motion $\dot{\phi}_i = 1$ and their dispersion relation is $p = E$ in appropriate units (the canonical momentum, which includes contributions from the Chern--Simons term on the worldvolume, is null).
Hence, we will continue to say, loosely, that they move ``at the speed of light.''}
Finally, for the non-extremal solutions, to keep the total energy of the excitations on the branes fixed, we will need to scale the non-extremality parameter $\mu$ appropriately.
With these $\epsilon$ scalings of $\ell_s$ and the coordinates, accompanied by appropriate $\epsilon$ scalings of the parameters of the solution, every term in the metric and the matter fields can be expanded in powers of $\epsilon$.
If the leading terms in this expansion give a homogeneous rescaling of the metric ($g_{\mu\nu} \to \epsilon^\alpha g^0_{\mu\nu}$), then $g^0_{\mu\nu}$ (accompanied by the leading terms in the $\epsilon$ expansion of the matter fields) must itself provide a solution to Einstein's equations.
This isolates the geometry dual to the decoupled D-brane gauge theory.
Below we will carry out this decoupling procedure separately for one-, two-, and three-charge superstars.

\subsection{One R-charge}
Set $q_2 = q_3 = 0$, and denote $q_1$ for simplicity by $q$.
Then the D-branes in (\ref{eq:nonextremal2}) are distributed along the angle $\theta_1$.
We let
\begin{equation}
\ell_s \to \epsilon \, \ell_s
\Label{lsscaling}
\end{equation}
and $L$  and $q$ will scale with their powers of $\ell_s$ so that
$ L \to \epsilon \, L \ {\rm and} \ q \to \epsilon^2 \, q \, . $
Strings stretched along $r$ and $\theta_1$ between branes at the angle $\theta_0$ and a probe giant graviton at $(r,\theta_0-\theta)$ will have a mass
\begin{equation}
m^2 = {{r}^2 + L^2 \, {\theta}^2 \over \ell_s^4} \,.
\end{equation}
To keep these masses fixed\footnote{
Here we are taking a probe approximation where we are computing the length of strings in the background AdS spacetime before the branes backreact to produce the superstar geometry.
This is because the backreacted geometry should be thought of as being {\it equivalent} to the physics of the light open strings in the probe treatment.}
we scale
\begin{equation}
r \to \epsilon^2 \, r \,; \qquad \theta_1 \equiv \theta_0 - \theta \to  \theta_0 - \epsilon \, \theta \,.
\Label{onechargethetascaling}
\end{equation}
It is also convenient to introduce a dimensionless time
$
\hat{t} =  t/L
$
which is held fixed in the scaling limit so that
\begin{equation}
t = L \, \hat{t} \to \epsilon \, L \,  \hat{t} \,.
\end{equation}
At any fixed $\theta_0$, the giant gravitons are moving along the $\phi_1$ circle at the speed of light.
Hence to keep the lengths of strings stretched to these branes fixed, it is necessary to focus into this lightlike direction.
In terms of the differential
$
d\chi = d\phi_1 - d\hat{t}
$,
it turns out that the appropriate rate of focusing is
\begin{equation}
d\chi \to \epsilon \, d\chi
\Label{chiscaling}
\end{equation}
in order to get a homogeneous scaling of the metric.

We must also keep fixed the energy above extremality associated to the branes at the fixed angle $\theta_0$.
To do this recall that non-extremality adds a mass to the black hole (\ref{mass})
\begin{equation}
M \sim {\mu \over G_5} \,.
\end{equation}
This mass integrates contributions from branes distributed over all angles $\theta$ on the $S^5$.
Hence we can estimate that the contribution to the mass from excitations on branes at any fixed angle is
\begin{equation}
M_\theta \sim {\mu \over G_5 L} \,.
\end{equation}
Since $G_5 \propto \ell_s^3$, to keep $M_\theta$ fixed in the limit (\ref{lsscaling}), we should take
\begin{equation}
\mu \to \epsilon^4 \, \mu \,.
\end{equation}
Scaling $\ell_s$, $q$, $L$, $t$, $r$, $\theta_1$, $\chi$, and $\mu$ in this way, we find that
\begin{equation}
  H_1 \sim \frac{q}{r^2} \,,
  \qquad f(r) \sim 1 + {q \over L^2} - \frac{\mu}{r^2} \,,
  \qquad \gamma \sim \frac{q}{r^2}\,\sin^2\theta_0 \,,
\end{equation}
and the metric scales {\it homogeneously} to leading order in $\epsilon$ as $\epsilon^3$.
The five-form field strength (\ref{flux}) takes the form
\bea
dB^{(4)} &=& \frac{2 r q}{L} \sin^2\theta_0\ dt\wedge\ dr\ \wedge d^3\Omega + 2 L^2 q \sin\theta_0\cos\theta_0\ d\chi_1\wedge d\theta\wedge d^3\Omega \,, \nonumber \\
\star dB^{(4)} &=& 2 L^4 \cos\theta_0\ d\chi_1\wedge d\theta\wedge d^3\widetilde\Omega + \frac{2 L^3 r}{q} \sin^2\theta_0\ dt\wedge dr\wedge d^3\widetilde\Omega \label{flux2}
\eea
and scales homogeneously as $\epsilon^6$.
Thus, both the Ricci scalar as well as the square of the five-form scale as $\epsilon^{-3}$, and the equations of motion will scale homogeneously as well.

As we have explained above, the leading terms in the metric give the geometry dual to a decoupled field theory on the giant gravitons located at $\theta_1 = \theta_0$.
In terms of the coordinate
\begin{equation}
z= {r \over \sqrt{q}}
\end{equation}
the metric is
\begin{equation}
ds^2 = \sin\theta_0 \left\{ z\left[ -f \, dt^2 + q \, ds^2_{S^3} + L^2 \, ds^2_{\tilde{S}^3} \right]
+ {1\over z} \left[ {q \over f} \, dz^2 + L^2 \, d\theta^2 + {L^2 \over \tan^2\theta_0} \, d\chi^2 \right]
\right\} \,,
\Label{1decoupleTheta}
\end{equation}
with
\begin{equation}
f = 1 +{q \over L^2} - { (\mu/q) \over z^2} \,. \Label{feqn1}
\end{equation}
To determine what field theory is dual to this geometry we have to compute how many D-branes are at the angle $\theta_0$ and what size they are.

Following \cite{MST}, giant gravitons can carry an R-charge $1 \leq J \leq N$, and in the probe brane approximation are located on the $S^5$ in $\ads{5} \times S^5$ at the angle
\begin{equation}
\sin\theta_1 = \sqrt{{J \over N}}
\Label{giantangle}
\end{equation}
and have a size
\begin{equation}
r = L \, \sin\theta_1 = L \, \sqrt{{J \over N}} \,.
\Label{giantsize2}
\end{equation}
This suggests that in the decoupling limit that we have been discussing, there is only a single giant graviton present at each $\theta_0$ and hence a $U(1)$ theory is dual to the geometry (\ref{1decoupleTheta}).
To test whether this is the case we should ask whether the strings stretched between a D-brane of R-charge $J$ and another of R-charge $J + k$ can remain light in our decoupling limit.
Again taking the probe approximation,\footnote{
{\em I.e.}, we do not include the backreacted metric of giant gravitons, and treat them as probes distributed on $S^5$.}
the mass of a string stretched between such branes will be
\begin{equation}
m = {L \over \ell_s^2} \, \Delta \theta
= {L \over \ell_s^2} \left[ \sin^{-1}\sqrt{{J+k \over N}} - \sin^{-1}\sqrt{{J \over N}} \right] \,.
\Label{stretched1}
\end{equation}
If we assume that $k\ll J\sim N$,
\begin{equation}
  \Delta\theta = \frac{k}{N\,\sin 2\theta_0} \,.
 \Label{k-theta}
\end{equation}
Since $L \propto \ell_s$,  the decoupling limit that we have been
discussing, $m \to m/\epsilon$ for fixed $k$, which diverges as
$\epsilon \to 0$. Thus it would appear that the branes of
different R-charges decouple from each other, and that the
geometry (\ref{1decoupleTheta}) is dual to a $U(1)$ gauge theory
on a sphere of size (\ref{giantsize2}).   An alternative way to
see this is that, taking $J \sim N$, the stretched open string
mass is held fixed in our scaling limit if we scale
$k\to\epsilon\,k$.  (This correctly reproduces the scaling of
$\Delta \theta$ in (\ref{onechargethetascaling}) via
(\ref{k-theta}).)  Thus it seems that in this limit, only the
strings on the brane located precisely at $\theta_1 = \theta_0$
(i.e. $k=0$) are massless, though to be sure of this conclusion we
would need to understand the wavefunctions of giant gravitons of
the $S^5$ and how they overlap.\footnote{In the regime where the
strings stretching between adjacent stacks of branes are light it
may as well be useful to consider a DLCQ limit of type IIB string
theory \cite{shahin}.}


\paragraph{Semiclassical limit and scaling $N$: }
Our previous analysis kept both $g_s$ and $N$ fixed as $\ell_s\to \epsilon\,\ell_s$, so that the AdS scale $L$ was also scaled to zero in the decoupling limit.
However, the semiclassical limit in which the geometry is reliable also requires that $N \to \infty$ to keep the AdS scale $L$ fixed (see, {\em e.g.}, \cite{juanAdS, babel1}).
We will keep the string coupling $g_s$ fixed and small, so that we can trust the tree-level supergravity approximation.
Thus, requiring $L$ to be fixed, determines the scaling in $N$:
\begin{equation}
N \to {N \over \epsilon^4} \,.
\Label{Nscaling}
\end{equation}
The mass of the stretched open strings \eqref{stretched1} in the regime $k\ll J\sim N$ behaves like
\begin{equation}
m \approx {L \over \ell_s^2} \, \frac{k}{N} \,.
\end{equation}
To keep $m$ fixed as $\ell_s \to \epsilon \, \ell_s$, we need
\begin{equation}
  k \to {k \over \epsilon^2} \,.
\Label{kscaling}
\end{equation}
Using \eqref{k-theta}, this scaling is consistent with a quadratic scaling in $\theta$.
Indeed, to get a uniform $\epsilon$ scaling of the metric we can take\footnote{
There is a one-parameter family of rescalings under which $r\to \epsilon^2 \, r$, $\mu\to\epsilon^4 \, \mu$, $\theta_1\to \theta_0-\epsilon^a \, \theta$, $\chi\to\epsilon^a \chi$, $t\to\epsilon^{2-a} \, t$, $L\to\epsilon^{2-a} \, L$ and
$q\to\epsilon^{4-2a} \, q$.
These all lead to the same metric (scaling as $\epsilon^{4-a}$) and five-form (scaling as $\epsilon^{8-2a}$).
This corresponds to a scaling $k\to \epsilon^{4-3a}\,k$.}
\begin{eqnarray}
&& L\,,g_s = {\rm fixed} \,; \qquad q = {\rm fixed} \,; \qquad t = L \, \hat{t} = {\rm fixed} \,; \nonumber \\
&& r \to \epsilon^2 \, r \,; \qquad \mu \to \epsilon^4 \mu \,; 
\qquad \theta_1 \to \theta_0 - \epsilon^2 \, \theta \,; \qquad d\chi \to \epsilon^2 d\chi \,,
\Label{q1scaling}
\end{eqnarray}
resulting in exactly the same decoupled metric and flux as in (\ref{1decoupleTheta}) and (\ref{flux2}).
With these scalings, including (\ref{Nscaling}), we can use (\ref{giantsize}) and (\ref{giantdist}) to estimate that the number of branes located in the vicinity of $\theta_0$ is
\begin{equation}
K = N_1 \sin2\theta_0 \, \Delta\theta
\end{equation}
where $\Delta\theta$ is the extent in $\theta$ over which open strings have fixed masses.   Then adopting the scaling (\ref{kscaling}) and using (\ref{k-theta}) we find that
\begin{equation}
K = {N_1 \over N} \, k \, ,
\end{equation}
where $k$ is a constant that must be determined by consideration of the wavefunction of giant gravitons on $S^5$.   This implies that  the geometry (\ref{1decoupleTheta}) is dual to a $U(K)$ gauge theory on a sphere of radius given by (\ref{giantsize}).

\paragraph{Interpretation of the metric: }
We can try to relate (\ref{1decoupleTheta}) to more familiar metrics.   Suppose we write down the near-horizon metric for $k$ D3-branes wrapping a three-sphere, with transversal space ${\cal C}\times T^2$, where ${\cal C}$ is a cone over $S^3$.  Also assume that the D3-branes are smeared over the transverse two-torus.  If we ignore the fact that this system is not a solution of type IIB supergravity, and if we employ the usual harmonic functions to describe D3-brane metrics, the resulting metric takes the same form as (\ref{1decoupleTheta}).  This suggests that (\ref{1decoupleTheta}) is dual to a suitable $U(K)$ gauge theory.  It is not clear whether the natural description is in terms of quantum mechanics, a four-dimensional gauge theory or even a six-dimensional theory, as the presence of the transverse two-torus might suggest.   We leave a more precise dual description of (\ref{1decoupleTheta}) for future work.

\paragraph{Black hole entropy: }
The entropy of the near-extremal single R-charged $\ads{5}$ black hole is:
\begin{equation}
  S_{\text{near-extremal}} = \pi\,\sqrt{N_1}\,N^{3/2}\,\left(\frac{r_h}{L}\right)^2 \,.
\end{equation}
Using the Bekenstein--Hawking entropy formula for our decoupled metrics, we obtain that the entropy stored in the focusing region located at $\theta_0$ is:
\begin{equation}
  S_{\text{strip}} =  \pi\,\sqrt{n_1}\,N^{3/2}\,\left(\frac{r_h}{L}\right)^2\,\left[4\sin^3\theta_0\,\cos\theta_0\,\delta\theta_0\right] \,.
\end{equation}
This result can be interpreted as the entropy density in that location:
it is equal to the total entropy $S_{\text{near-extremal}}$ times a fraction less than one that is regulated by the size of the focused area $\delta\theta_0$.
By integrating this density over $\theta_0 \in [0,\,\pi/2]$ and using the identity
\begin{equation}
  4 \int_0^{\pi/2} \delta\theta_0 \,\sin^3\theta_0\,\cos\theta_0 = 1 \,,
\end{equation}
we recover the total entropy of the original $\ads{5}$ black hole:
\begin{equation}
  \int S_{\text{strip}} = S_{\text{near-extremal}} \,.
\end{equation}
The precise interpretation of this entropy computation is not quite clear, but our results above suggested that each decoupled ``strip'' is dual to a $U(N_1 \sin2\theta_1 \, \Delta \theta) = U(K)$ theory whose entropies then add up to give the total degeneracy.
This would imply  that the entire near-horizon geometry is dual to a
\begin{equation}
\prod_{\theta_i} U(K)
\end{equation}
gauge theory where the product runs over  a set of discrete angles at which the giant gravitons are found. We might think about this as the Coulomb branch of the theory on the $U(N_1)$ giant gravitons that are sourcing the superstar.
To justify this, recall that while we have been thinking about the superstar in terms of giant gravitons (spherical branes on $S^3$), but we could equally well have thought about the spacetime in terms of dual giant gravitons (spherical branes that expand into $\ads{5}$).
In terms of the latter, the $SU(N)$ Yang--Mills theory dual to the spacetime is in the Coulomb branch and is Higgsed down to a product of $U(k)$ factors depending on the numbers of coincident branes \cite{IH}.
Our results suggest that we can similarly take the perspective that the near-horizon geometries of the single-charge superstars have a dual description in the Coulomb branch of the gauge theory on the $N_1$ giant gravitons that source the spacetime.
To treat this Coulomb branch properly we would need to work out the wavefunction of the giant gravitons on $S^5$ to see how these wavefunctions overlap and splice together in strips to cover the $S^5$.
We have not attempted to do this, but some evidence for this perspective is given below.

\paragraph{New dualities: }
We can now show that the metric (\ref{1decoupleTheta}) appears as a decoupling limit of two {\it different} asymptotic geometries.
To see this, write (\ref{totalgiants}) as
\begin{equation}
q = {N^\prime \over N} L^2 \,; \qquad L = \ell_s \, (4\pi g_s N)^{1/4} \,,
\end{equation}
where $N^\prime$ is the total number of giant gravitons in the solution.
Then define the new variables
\begin{equation}
q^\prime = {N \over N^\prime} L^{\prime 2} \,; \qquad
L^\prime = \ell_s \, (4\pi g_s N^\prime)^{1/4} \,.
\end{equation}
In terms of $\hat{q}^{\prime} = q^{\prime}/L^{\prime 2}$, we can define new parameters and coordinates
\bea
r' &=& r \left( \frac{q}{L^2} \right)^{-1/4} \,, \\
\mu' &=& \mu \left( \frac{q}{L^2} \right)^{-3/2} \,,
\eea
with no change in either $\hat{t}$ or $d\chi$.
Notice that $L'^2=L\sqrt{q}$.
In terms of these variables and coordinates, the metric (\ref{1decoupleTheta}) becomes
\begin{equation}
ds^2 = \sin\theta_0 \left\{ z^\prime \left[ -f^\prime \, dt^{\prime 2} + L^{\prime 2} \, ds^2_{S^3} + q^\prime \, ds^2_{\tilde{S}^3} \right]
+ {1\over z^\prime} \left[ {q^\prime \over f^\prime} \, dz^{\prime 2} + L^{\prime 2} \, d\theta^2 + {L^{\prime 2} \over \tan^2\theta_0} \, d\chi^{\prime 2} \right]
\right\} \,,
\Label{duality1}
\end{equation}
where $z^\prime$ and $f^\prime$ found from $z$ and $f$ by
replacing unprimed with primed variables. The form of the metric
is exactly the same as (\ref{1decoupleTheta}) except that the two
$S^3$ factors have been exchanged. This is essentially an exchange
between giant gravitons and dual giant gravitons and tells us that
the near-brane physics can be described equivalently either in
terms of the $U(N)$ Yang--Mills theory dual to the entire
spacetime, or in terms of the theory on the giant gravitons.
Indeed, in the form (\ref{duality1}), the metric could be extended
to an asymptotically $\ads{5}$ geometry with scale $N^\prime$
rather than $N$. This tells us that the deep infrared physics of
$U(N)$ gauge theory with R-charge $N^\prime$ is equivalent to the
deep infrared of $U(N^\prime)$ gauge theory with R-charge $N$ as
suggested in \cite{vijayasad}.

So far, we have focused on the invariance of the metric, but it is
easy to check the same is true for the five-form flux, in which we
are exchanging $dB_4$ with $\star\,dB_4$, due to the exchange
between the two three-spheres. We note also that under these
definitions, the parameters in the leading terms of the
gravitational entropy \eref{eq:sbh1} and field theoretic entropy
\eref{fine} become primed.

The above infrared duality is not unique.
Indeed, we can also exchange the rank of the gauge group $N$ with the number of giants $N^\prime$ (exchanging both $S^3$ factors) by rescaling the string coupling, while keeping the AdS scale fixed \cite{oldnote}:
\begin{equation}
  L^\prime = L \,, \quad q^{\prime} = \frac{L^4}{q} \,\,\Longrightarrow \,\, g_s^\prime = g_s\,\hat{q}\,\,\Longrightarrow \,\, G_{10}^\prime = G_{10}\,\hat{q}^2 \,,
\end{equation}
where we assumed the string scale $\ell_s$ is kept fixed.
The rescaling of the ten-dimensional Newton constant means that the ten-dimensional metric scales like
 \begin{equation}
ds^2(q^\prime, \mu^\prime) = \sqrt{\hat{q}}\,\left( ds^2(q,\mu), S^3 \leftrightarrow \tilde{S}^3 \right) \,,
\end{equation}
which is indeed the case when accompanied by the changes \cite{oldnote}:
\begin{equation}
  z^\prime = z\,\sqrt{\hat{q}} \,, \quad \mu^\prime = \frac{1}{\hat{q}^2}\,\mu \,.
\end{equation}
It is reassuring that the five-form flux behaves as in the previous duality, but with a scaling of $\hat{q}$, {\em i.e.},
$\left(dB_4 + \star\,dB_4\right)^\prime = \hat{q}\,\left(dB_4 + \star\,dB_4\right)$,
which is in agreement with the scaling of the Newton constant.
Thus, this second transformation is also a symmetry of the infrared physics, and it is still weakly coupled, since the number of giants is bounded from above by $N$.

The rescaled metric extends to infinity to give an asymptotically
$\ads{5}$ metric with the same AdS scale, but a $U(N^\prime)$ dual
gauge group and a rescaled coupling. This is precisely the duality
between the infrared of $U(N)$ gauge theory with R-charge
$N^\prime$ and $U(N^\prime)$ gauge theory with R-charge $N$ that
was suggested in \cite{vijayasad}.\footnote{ This is related to a
$\BZ_2$ symmetry that relates $U(N)$ Chern-- Simons theories at
level $N'$ to $U(N')$ Chern--Simons theories at level $N$
\cite{iks,ms}.}

The reason that we apparently have more than one infrared duality
is identical to the reason that we had several decoupling limits:
in the original metric (\ref{eq:nonextremal2}) we can rescale
$t\to L \, \hat{t}$, $q_i\to L^2 \, \hat{q}_i$, $r\to L \,
\hat{r}$, and $\mu \to L^2 \, \hat{\mu}$. The resulting metric is
then independent of $L$, except for an overall factor of $L^2$ in
front of the metric. Clearly, we can rescale $L$ any way we want,
while keeping the hatted variables fixed, and the metric will
still be a solution of the equations of motion. It is easy to see
that both the decoupling limits as well as the two infrared
dualities presented above are related to each other by such a
rescaling of $L$.

\paragraph{Scaling to $\theta_1 = 0, \pi/2$: }
The scaling limit described above produced the geometry dual to the decoupled theory on branes at any finite angle $0<\theta_1<\pi/2$.
Maximal and minimal sized giant gravitons, {\em i.e.}, $\theta_1=\pi/2$, and $\theta_1=0$, respectively, require a different analysis.

Let us consider the maximal sized giant graviton at $\theta_1 = \pi/2$.
We do obtain an homogeneous scaling of the metric $(\epsilon^2)$ in the limit
\begin{equation}
\theta_1 \equiv \frac{\pi}{2} - \theta \to \frac{\pi}{2} - \epsilon^2 \, \theta \,; \quad
d\chi \,\,\, {\rm fixed} \,,
\end{equation}
with all other scalings as in (\ref{q1scaling}):
\begin{equation}
ds^2 =  z\left[ -f \, dt^2 + q \, ds^2_{S^3} + L^2 \, ds^2_{\tilde{S}^3} \right]
+ {1\over z} \left[ {q \over f} \, dz^2 + L^2 \, d\theta^2 + L^2  \, \theta^2 \, d\chi^2 \right] \,,
\Label{1decouplePi2}
\end{equation}
with $f$ as in (\ref{feqn1}).
This geometry is dual to the low energy theory on the maximal giant graviton.
Likewise, for the vanishing size ``giant,''  we can consider the limit
\begin{equation}
\theta_1 \to \epsilon^2 \, \theta \,; \quad
d\chi \to \epsilon^4 \, d\chi \,,
\end{equation}
with all other scalings as above.
This gives the decoupled metric
\begin{equation}
ds^2 = \sqrt{1 +  {\theta^2 \over z^2}} \left\{ z^2 \left[ -f \, dt^2 + q \, ds^2_{S^3} + {L^2 \, \theta^2 \over \theta^2 + z^2} \, ds^2_{\tilde{S}^3} \right]
+ \left[ {q \over f} \, dz^2 + L^2 \, d\theta^2 + {L^2 \over z^2 + \theta^2} \, d\chi^2 \right]
\right\} \,,
\Label{1decouple0}
\end{equation}
with $f$ as in (\ref{feqn1}).
However, because the giant gravitons at $\theta_1 \to 0$ are too small to be D-branes (they are understood as regular gravitons), it is unclear what dual field theory would describe this geometry.

\subsection{Two R-charges}

\label{2charge}

The two-charge superstar contains two species of giant gravitons that are rotating in different directions on the $S^5$.
It was originally proposed by Gubser and Heckman \cite{gubserheckman} that strings living on the intersection of the two species of giant gravitons could account for the entropy of the two-charge superstar.
We will provide further substance for this idea (and discuss various subtleties) by displaying a decoupling limit in which a metric with a warped $\ads{3}$ metric appears.
Unlike the single-charge case, we have not developed a systematic argument showing that it is precisely the open strings on the giant graviton intersections that are kept light in our decoupling limit.
Nevertheless, the form of the decoupled metric, and its dependence on the number of giant graviton intersections, will provide evidence that this is indeed the case.

In the two-charge case we take $q_1 = 0$ in (\ref{eq:nonextremal2}) so that the D-branes are rotating in the $\phi_2$ and $\phi_3$ directions and have sizes \cite{superstar}
\begin{equation}
\rho_2 =  L \sqrt{1 - \mu_2^2} ~~~~;~~~~ \rho_3 =  L \sqrt{1 - \mu_3^2} \,,
\end{equation}
where as before
\be
\mu_1 = \cos\theta_1 \,; \qquad \mu_2 = \sin\theta_1 \, \cos\theta_2 \,; \qquad \mu_3 = \sin\theta_1 \,\sin\theta_2 \,.
\ee
They mutually intersect on the $\phi_1$ circle with a size $L \, \cos\theta_1$.
The dependence of the two-charge entropy (\ref{2chargeentropy}) on the product of charges suggests that the near-extremal entropy is dominated by strings running between the two species of branes and thus localized on the brane intersections.

As before we will take
\be
\ell_s \to \epsilon \, \ell_s \,
\ee
with
\be
L = {\rm fixed} ~~~~ \Longrightarrow ~~~~ N \to {1 \over \epsilon^4} N \,.
\ee
Since we are interested in two species of giant gravitons, both of which are moving on the $S^5$ at the speed of light, to keep the masses of stretched strings fixed we will need to focus in on both $d\theta_1$ and $d\theta_2$ as well as on the null directions
\begin{equation}
d\chi_2 = d\phi_2 - d\hat{t} \,; \qquad d\chi_3 = d\phi_3 - d\hat{t} \,.
\end{equation}
The scalings of $r$ and $t$ are the same as in the one-charge case.
Thus we take
\begin{eqnarray}
&&
\theta_1 \to \theta_1^0 - \epsilon \, \theta_1 \,; \qquad
\theta_2 \to \theta_2^0 - \epsilon \, \theta_2
~~~~ \Longrightarrow ~~~~ d\mu_{i} \to \epsilon \, d\mu_i \,,
\label{sca1}
\\
&& \qquad \qquad \qquad
d\chi_2 \to \epsilon \, d\chi_2 \,; \qquad d\chi_3 \to \epsilon \, d\chi_3 \,; \\
&& \qquad \qquad \qquad
r \to \epsilon^2 \, r \,; \qquad t \equiv L \, \hat{t} = {\rm fixed} \,; \\
&& \qquad \qquad \qquad
q_i \to \epsilon^2 \, q_i \,; \qquad \mu \to \epsilon^4 \, \mu \,. \label{sca4}
\end{eqnarray}
Here $0 \leq \theta_i^0 \leq \pi/2$ are fixed values of $\theta_i$ and
\be
d\mu_i = \left[ {\partial\mu_i \over \partial \theta_1} \right]_{\theta_1^0,\theta_2^0} \, d\theta_1 + \left[ {\partial\mu_i \over \partial\theta_2} \right]_{\theta_1^0,\theta_2^0} \, d\theta_2 \,.
 \Label{mutheta}
\ee
The scaling of $q_i$ is fixed by its proportionality to $\ell_s^2$, and the scaling of the non-extremality parameter $\mu$ is fixed as in the single-charge case.
With these scalings, the leading terms terms in the metric scale as $\epsilon^2$ and give
\begin{multline}
ds^2 = \mu_1^0 \left[ - {r^2 \over \sqrt{q_2 q_3}} \, f \, dt^2 + {\sqrt{q_2 q_3} \over r^2}\frac{1}{f}
 \, dr^2 + {L^2 \, r^2 \over \sqrt{q_2 q_3}} \, d\phi_1^2 +  \sqrt{q_2 q_3} \, ds^2_{S^3} \right]  \\
+ {L^2 \over \mu_1^0 \sqrt{q_2 q_3}} \left[ \sum_{i=2,3} q_i (d\mu_i^2 + (\mu_i^{0})^2 \, d\chi_i^2) \right] \,,
\label{2scalemet1}
\end{multline}
with
\be
f = 1 - {\mu \over r^2} + {q_2 q_3 \over L^2 r^2} \,.
\ee
Likewise the leading terms in the flux are
\begin{equation}
  B_4 = -L^2\,\sum_{i=2}^3 q_i\,\mu_i^2\,\left(d\phi_i - d\hat{t}\right)\wedge d^3\Omega \,,
  \Label{2scaleflux}
\end{equation}
giving rise to a five-form flux which scales as $\epsilon^4$:
\begin{equation}
  \frac{F_5}{L^2} = 2\sum_{i=2}^3 q_i\,\mu^0_i\,d\mu_i\wedge (d\phi_i-d\hat{t})\wedge d^3\Omega -2 \frac{L^2 r}{q_2 q_3} \sum_{i=2,3} (q_i \mu_i^0\ d\hat{t}\wedge dr\wedge d\phi_1\wedge d\mu_i\wedge d\chi_i) \,.
\end{equation}
Taken together, (\ref{2scalemet1}) and (\ref{2scaleflux}) give a
solution to Einstein's equations which should be dual to the
theory of open strings localized on the intersection of the giant
gravitons at the angular location $\theta_1^0, \theta_2^0$. One
also sees this by analogy with the D1-D5 system where the 5-5
strings and the 1-1 strings become non-dynamical while the 1-5
strings, which live on the intersection circle, continue to
fluctuate. In that case, the circle associated to the 1-5 strings
shrinks as $r \to 0$ in the near-horizon limit, just as the
$\phi_1$ circle shrinks here.

\paragraph{Scaling to $\theta_1 = 0$: }
Above we took a scaling limit that focused in to a generic
location on $S^5$. At the special angle $\theta_1 = 0$, the giant
gravitons are stationary and of maximal size \cite{superstar}.
Because of this, the appropriate scaling limit is
\be
\theta_1 \to \epsilon \, \theta_1 \,; \qquad \theta_2, \, d\chi_2, \, d\chi_3 = {\rm fixed} \,,
\ee
with all other variables scaling as above.
The leading terms in the metric again scale as $\epsilon^2$ and decoupled metric is
\begin{multline}
ds^2 = \left[ - {r^2 \over \sqrt{q_2 q_3}} \, f \, dt^2 + {\sqrt{q_2 q_3} \over r^2}\frac{1}{f}
 \, dr^2 + {L^2 \, r^2 \over \sqrt{q_2 q_3}} \, d\phi_1^2 +  \sqrt{q_2 q_3} \, ds^2_{S^3} \right]  \\
+ {L^2 \over \sqrt{q_2 q_3}} \left[ q_2 (d\mu_2^2 + \theta_1^2 \, \cos^2\theta_2\, d\chi_2^2) +
 (d\mu_3^2 + \theta_1^2 \, \sin^2\theta_2\, d\chi_3^2)
 \right] \,.
\end{multline}

\paragraph{Interpretation of the decoupled solution: }
To interpret the decoupled metric (\ref{2scalemet1}) we define
\be
\ell =  (q_2 q_3)^{1/4} \,; \qquad \tilde{\ell} = \ell \,
(\mu_1^0)^{1/2} \,; \qquad \tilde{t} = t \, {\tilde{\ell} \over L}
\,; \qquad \tilde{r} = r \, {L \over \tilde{\ell}} \, \mu_1^0 \,.
\ee
In terms of these variables we can write
\be
\tilde{f} = {\tilde{r}^2 \over \tilde{\ell}^2 } - M_3 \,; \qquad
M_3 = {\mu \, L^2 \over q_2 q_3} - 1 \,,
\ee
in terms of which (\ref{2scalemet1}) becomes
\be
ds^2 = \left[ - \tilde{f} \, d\tilde{t}^2 + {1\over \tilde{f} } \,
d\tilde{r}^2 + \tilde{r}^2 \, d\phi^2 + \tilde{\ell}^2 \,
ds^2_{S^3} \right] + {L^2 \over \tilde{\ell}^2} \left[
\sum_{i=2,3} q_i (d\mu_i^2 + (\mu_i^0)^2  \, d\chi_i^2 ) \right] \,.
\Label{ads3met}
\ee
In the extremal limit
\be
\mu = 0 ~~~~\Longrightarrow~~~~ M_3 = -1 \,,
\ee
and the terms in the first square brackets are precisely the metric of $\ads{3} \times S^3$ where both factors have a scale $\tilde{\ell}$.
The second square brackets enclose a flat metric --- identifying the $\xi_i$ and the $\mu_i$ (or $\theta_i$) periodically gives a $T^4$.
Thus we have found a new class of $\ads{3} \times S^3 \times T^4$ solutions in type IIB string theory that are supported entirely by five-form flux.
If we instead wrote the solution in terms of the scale
\be
\ell = (q_2 q_3)^{1/4}
\ee
and treated $\sin\theta_i^0$ and $\cos\theta_i^0$ as functions of $\theta_i$ rather than fixed numbers, these solutions resemble the $1/8$-supersymmetric warped $\AdS{3}$ metrics in type IIB supergravity that were studied in \cite{kim1,kim2}.
In the range
\be
0 < \mu < {q_2 q_3 \over L^2} = \mu_c \,; \qquad -1 < M_3 < 0
\ee
the metric (\ref{ads3met}) precisely describes a conical defect in the $\ads{3}$ factor.
This explains the curious fact that for the two-charge superstar geometry in five dimensions the horizon vanishes for $0 < \mu < \mu_c$ giving a geometry with a naked singularity.
We now see that in the decoupled geometry there is a conical defect, suggesting that these geometries are all resolved in string theory and are replaced by analogues of the family of non-singular, horizon-free solutions in $\ads{3}$ described in \cite{conical,maldmaoz,mathurlunin,maldmaozlunin}.
Finally, when
 \be
 \mu \geq \mu_c \,; \qquad M_3 \geq 0
 \ee
the $\ads{3}$ factor in (\ref{ads3met}) contains a $J=0$ BTZ black hole with a horizon radius and entropy
\be
\tilde{r}_h = \tilde{\ell} \sqrt{M_3} \,; \qquad
  S_{BTZ} = {A_{BTZ} \over 4 G_3} = {2\pi r_h \over 4 G_3} = {\pi \tilde{\ell} \sqrt{M_3} \over 2 \, G_3} \,.
\ee
Interestingly, two T-dualities on the $T^4$ factor of the metric converts the Ramond--Ramond four-form potential into a two-form potential.
Thus, this T-dual framework will involve the near-horizon limit of a D1-D5 system and the framework of \cite{conical,maldmaoz,mathurlunin,maldmaozlunin} can be applied directly.

\paragraph{Black hole entropy: }
The decoupled geometry (\ref{ads3met}) will have a dual description in a $(1+1)$-dimensional CFT which is associated to the theory on the intersection of the D-branes in the original superstar solution.
Following \cite{adsentropies}, we can relate the left- and right-moving Hamiltonians ($L_0, \bar{L}_0$) and the central charge of the CFT to the geometries (\ref{ads3met}) as
\be
L_0 = \bar{L}_0 = {M_3 \, \tilde{\ell} \over 16 \, G_3} \,; \qquad c = {3 \, \tilde{\ell} \over 2 \, G_3} \,,
\ee
where $G_3$ is the effective three-dimensional Newton constant after compactifying on $S^3 \times T^4$.
The entropy is then exactly reproduced by the Cardy formula applied to the $(1+1)$-dimensional CFT:
 \be
 S_{CFT} = 2\pi\sqrt{c \,L_0/6} + 2\pi \sqrt{c \, \bar{L}_0/6} = S_{BTZ} \, .
 \ee
To relate this microscopic derivation of the black hole entropy in the decoupled theory to the original superstar we must first relate $G_3$ to $G_{10} = 8\pi^6 (g_s^2 \ell_s^8)$:
 \be
 {1 \over G_3} =  {V_7 \over G_{10}}
 = \left( {N^2 \, \tilde{\ell}^3 \over L^4} \right) \left( {4 \over \pi^2} \right) \left(  {\mu_2^0 \, \mu_3^0 \over (\mu_1^0)^2} \right) \, dV_4 \,,
\ee
where $V_7$ is volume of the seven dimensions transverse to the $\ads{3}$ and $dV_4$ is the product of the periodicities of the variables $\chi_i$ and $\mu_i$.
The central charge of the dual CFT is then
\be
c = \left( {6 \over \pi^2} \right) \left( {N^2 \, \ell^4 \over L^4 } \right) \left( \mu_2^0 \, \mu_3^0  \right) \, dV_4
= (N_2 \, \mu_2^0 \, d\mu_2) \,  (N_3 \, \mu_3^0 \, d\mu_3) \left( {6 \over \pi^2} \right) \, d\chi_2 \, d\chi_3 \,,
 \Label{c1}
\ee
where $N_i = q_i N/L^2$ is the total number of giants of species $i$.
Now given (\ref{giantdist}) for the distribution of giant gravitons in the superstar solution and (\ref{giantsize}) relating the size and position of the D-branes, it is easy to show that in our scaling limit
\be
dN_i = 2 (N q_i/L^2) \, \mu_i^0 \, d\mu_i = 2N_i \, \mu_i^0 \, d\mu_i
\ee
is the number of branes of species $i$ at the location $\theta_1^0,\theta_2^0$ on the $S^5$.
Thus the central charge is proportional to the product of the number of intersecting D-branes at the location on which we are focusing
\be
c= dN_2 \, dN_3 \,  \left( {6 \over 4 \, \pi^2} \right) \, d\chi_2 \, d\chi_3 \,,
 \Label{c2}
\ee
exactly as we should expect if the strings stretched between the intersecting branes are dominating the entropy.
The natural periodicity of $\chi_i$ is $2\pi$ giving a central charge
\be
c = 6 \, dN_2 \, dN_3 \,.
\Label{c3}
\ee
This is precisely the central charge of a supersymmetric non-linear sigma model on the target space
\be
(T^4)^{dN_2 \, dN_3} / S_{dN_2 \, dN_3}
\ee
suggesting that, locally, there are four bosons plus four fermions associated to the $(1+1)$-dimensional effective theory at the intersection of a pair of giant gravitons and that these describe fluctuations of the effective string into the $T^4$ factor of the geometry.

Nevertheless, once we glue the CFTs associated to all values of $\theta_i^0$ together, the na\"{\i}ve result for the total central charge becomes
\be c=6 \int dN_2 dN_3 = 6 N_2 N_3 \int d(\mu_2)^2 d(\mu_3)^2 = 3 N_2 N_3 \,, \Label{jjj1}
\ee
because $\mu_2^2,\mu_3^2$ have to obey $\mu_2^2 + \mu_3^2\leq 1$.
This value of the central charge has an appealing interpretation in terms of giant gravitons, as it is the central charge of a supersymmetric non-linear sigma model whose target space has complex dimension $N_2N_3$, exactly as expected for the CFT living on the intersection of the giant gravitons.
The latter will probe the moduli space of $1/4$-BPS giants, which according to \cite{mikhailov} can be described by intersections of a holomorphic polynomial $P(z_2,z_3)$ of degree $N_2$ in $z_2$ and degree $N_3$ in $z_3$ with the unit five-sphere in $\mathbb C^3$.
The complex dimension of the moduli space of such polynomials equals $N_2 N_3$ (up to $1/N$ corrections), in perfect agreement with the result (\ref{jjj1}) for the total central charge.

Turning back to the entropy, we would like to relate the entropy of the decoupled three-dimensional geometries to the entropy of the two-charge superstar in the near-extremal limit
\be
\mu- \mu_c \ll q_2, q_3 \ll L^2  \,.
\ee
This condition can be interpreted as stating that the number of giants (and the amount of non-extremality) should be much less than the number $N$ of D3-branes that gave rise to the original $\ads{5}$ geometry.
In this limit the five-dimensional entropy can be written as
\be
S_{{\rm two-charge}} = \pi {N^2 \ell^4 \over L^4} \sqrt{M_3}
\ee
where $M_3$ is as above.
Now recall that the the effective BTZ geometry that we have found arises as a focusing limit onto a particular value of $\theta_1, \theta_2$ and that the entropy on this strip of the horizon is accounted for by a CFT with the central charge (\ref{c1}).
The entropy associated with each such strip is
\be
dS_{BTZ} = {2 \over \pi} \, {N^2 \, \ell^4 \over L^4} \,  \sqrt{M_3}\, \mu^0_2 \, \mu^0_3 \, d\chi_2 \, d\chi_3 \, d\mu_2 \, d\mu_3 \,.
\Label{btzent}
\ee
Using the change of variables (\ref{mutheta}) it is easy to verify that
\be
d\mu_1\, d\mu_2 =
\cos\theta_1^0 \sin\theta_1^0 \, d\theta_1 \, d\theta_2 \,.
\ee
Recognizing $dS_{BTZ}$ as the entropy associated to a strip of the geometry at a given $\theta_1,\theta_2$ we would like to integrate (\ref{btzent}) over $\chi_i$ and $\theta_i$ which have ranges $0 \leq \chi_i \leq 2\pi$ and $0\leq \theta_i \leq \pi/2$.\footnote{
To verify these ranges, set all the charges and the non-extremality parameter to zero in which case the second line in the metric (\ref{eq:nonextremal2}) should parameterize an $S^5$.}
Putting these ranges in gives
\be
S = \int dS_{BTZ} = S_{{\rm two-charge}} \,,
\ee
precisely reproducing the entropy of the five-dimensional near-extremal black hole.

As in the single-charge case, the precise interpretation of this match is not clear, but it suggests that the near-horizon geometry of the two-charge superstar is dual to a product of CFTs, with central charges given by (\ref{c3}).
This product CFT would be regarded as the ``Coulomb branch'' of a CFT with a total central charge given by (\ref{jjj1}).
It would be nice to make this picture precise.

\paragraph{Exact solutions with warped $\ads{3}$: }
The metric (\ref{2scalemet1}) has an $\ads{3}$ factor and depends on the choice of angles $\theta_1^0,\theta_2^0$.
It therefore does not cover all global features of the giant graviton system.
It is an interesting question whether or not there are metrics that do contain such global information.
For example, one could imagine replacing $\mu_i^0$ by $\mu_i$ in (\ref{2scalemet1}).
The resulting metric has a warped $\ads{3}$ factor, and is reminiscent of the metrics in \cite{kim1,kim2}, but it is not a solution of the supergravity equations of motion and is therefore also not the result of a suitable decoupling limit.
Nevertheless, we find that in the case where the two-charges are equal, $q_2=q_3$, a small modification of (\ref{2scalemet1}) with $\mu_i^0$ replaced by $\mu_i$ is an exact solution of type IIB supergravity.
For the details we refer the reader to appendix~\ref{truesolution}.
This warped $\ads{3}$ metric possibly correctly captures the global features of the giant graviton system, but we leave a further study of its properties as well as possible extensions to $q_2\neq q_3$ to future work.

\paragraph{Adding a third R-charge perturbatively: }
It is worth noting that the third R-charge ($q_1$) corresponds to the quantum number associated to rotations in the $\phi_1$ direction as pointed out in \cite{gubserheckman}.
Hence, we might have expected to find a rotating BTZ metric rather than (\ref{ads3met}) in the three-charge case, at least in a perturbative expansion in $q_1$.
However, we have been unable to find a decoupling limit in which a such a rotating BTZ black hole appears.

\subsection{Three R-charges} \label{threecharge}

The three-charge AdS black hole contains three species of giant gravitons rotating in different directions of the $S^5$.
They mutually intersect on a point, and the number of such intersections is proportional to $N_1\,N_2\,N_3$, which also controls the entropy.
This suggests that the near-extremal entropy should be accounted for by the degrees of freedom living at such an intersection, and as such, it should be described by some (super)conformal quantum mechanics.

As before we will take
\begin{equation}
  \ell_s\to \epsilon\,\ell_s \,,
\end{equation}
but this time it is not possible to keep $L$ fixed while having a homogeneous scaling of the full ten-dimensional metric.
The decoupled metric is achieved by rescaling $t=L\,\hat{t}$ and taking
\begin{equation}
  L \to \frac{L}{\epsilon}\qquad \Longrightarrow \qquad N\to \frac{1}{\epsilon^8}\,N \,, \quad g_s\ \text{fixed} \,.
\end{equation}
We will require that $q_i$ and $r$ scale as in the two-charge case.
Hence the number of giants for any species scales like
\begin{equation}
  N_i \to \frac{1}{\epsilon^4}\,N_i\,,
\end{equation}
by consistency with $q_i/L^2=N_i/N$.

Since these black holes have three species of giant gravitons, all rotating at the speed of light, we will again need to focus on both the location on $S^5$ as well as on the null directions
\begin{equation}
  d\chi_i = d\phi_i - d\hat{t} \,, \,\,\,\,i=1,2,3 \,.
\end{equation}
The precise scaling that we take is:
\begin{eqnarray}
  d\mu_i\to \epsilon^2\,d\mu_i\,, \quad  & & \quad d\chi_i \to \epsilon^2\,d\chi_i\,, \\
  r\to \epsilon^2\,r\,, \quad & & \quad \hat{t} = \text{fixed}\,, \\
   q_i\to \epsilon^2\,q_i\,, \quad & & \quad \mu-\mu_c \to \epsilon^4\,\left(\mu-\mu_c\right)\,.
\end{eqnarray}
Above, $\mu_c = \left(q_1\,q_2+q_1\,q_3+q_2\,q_3\right)/L^2$ and it scales like $\mu_c\to \epsilon^6\,\mu_c$.
An important difference with previous scalings is that it is the difference $\mu-\mu_c$ that scales like the energy density $\epsilon^4$.
It is this double scaling limit and the fact that $\left(q_1\,q_2\,q_3\right)/(L^2\,r^4)$ remains fixed that allows us to retain the information about the horizon of the system in the decoupled metric.
This has an overall leading $\epsilon^2$ scaling and is given by
\begin{multline}
  ds^2 = - \frac{r^4}{\sqrt{q_1q_2q_3}}\,\sqrt{\hat{\gamma}}\,f\,L^2\,d\hat{t}^2 + \frac{\sqrt{q_1q_2q_3}}{r^2}\,\frac{\sqrt{\hat{\gamma}}}{f}\,dr^2 + \sqrt{q_1q_2q_3}\,\sqrt{\hat{\gamma}}\,ds^2_{S^3} \\
  + \frac{L^2}{\sqrt{q_1q_2q_3}\,\sqrt{\hat{\gamma}}}\,\sum_{i=1}^3 q_i\,\left(d\mu_i^2+ (\mu_i^0)^2\,d\chi_i^2\right)\,,
\end{multline}
where
\begin{equation}
  \hat{\gamma}=\sum_{i=1}^3 \frac{(\mu_i^0)^2}{q_i}\,, \quad
  f =  1 - \frac{\mu-\mu_c}{r^2} + \frac{q_1\,q_2\,q_3}{L^2\,r^4}\,.
\end{equation}

\paragraph{Interpretation of the decoupled metric:}
Since the metric is a direct product, let us focus on the two-dimensional metric in $\{\hat{t},\,r\}$.
Rewrite the metric in terms of
\begin{equation}
  t=L\,\hat{t}\,, \quad r^2=y\,,
\end{equation}
and observe that it is conformally flat
\begin{equation}
  ds^2 = e^{2\lambda}\,\left(-dt^2 + d\rho^2\right)\,,
\end{equation}
with conformal factor
\begin{equation}
  e^{2\lambda} = \frac{\sqrt{\hat{\gamma}}}{\sqrt{q_1\,q_2\,q_3}}\,\left(y^2 -(\mu-\mu_c)\,y + q_1\,q_2\,q_3/L^2\right)\,, \quad \frac{dy}{d\rho} = \frac{2}{\sqrt{\hat{\gamma}}}\,e^{2\lambda}\,.
\end{equation}
The curvature of this metric is controlled by the Laplacian of $\lambda(\rho)$:
\begin{equation}
  R_{\alpha\beta} = -\frac{4}{\sqrt{\hat{\gamma}\,q_1\,q_2\,q_3}}\,g_{\alpha\beta} \quad \Longrightarrow \quad
  R = -\frac{8}{\sqrt{\hat{\gamma}\,q_1\,q_2\,q_3}} \,,
\end{equation}
and we see that the decoupled metric has a two-dimensional spacetime with negative cosmological constant ($\ads{2}$), with AdS scale $\tilde{L}^2 \propto \sqrt{\hat{\gamma}\,q_1\,q_2\,q_3}$.

\paragraph{Black hole entropy: }
The near-extremal black hole entropy computed from the asymptotically $\ads{5}$ perspective equals (see (\ref{3ent}))
\begin{equation}
  S_{\text{near-extremal}} = \pi\,\sqrt{N}\,\sqrt{N_1\,N_2\,N_3}\,.
\end{equation}
Remarkably, in the near-extremal limit the leading term in the entropy is {\it independent} of $\mu$.
The structure of the entropy formula is reminiscent of the three-charge, finite area, extremal black holes in five dimensions originally studied in \cite{stromvafa}.
The entropy associated to the decoupled $\ads{2}$ metric is
 \begin{equation}
   S_{\text{strip}} =  \pi\,\sqrt{N}\,\sqrt{N_1\,N_2\,N_3}\,\left(8\,\mu_2^0\,\delta\mu_2\mu_3^0\,\delta\mu_3\right) \,,
 \end{equation}
whenever there exists a non-trivial horizon, which occurs when
 \begin{equation}
  \left(\mu-\mu_c\right)^2 > 4\,\frac{q_1q_2q_3}{L^2} \,.
\end{equation}
Once again, we can interpret the entropy of the decoupled metric as an entropy density associated with the location where we focused in.
Integrating this density with the induced metric discussed in the previous section reproduces the total entropy
\begin{equation}
  S_{\text{near-extremal}} = \int S_{\text{strip}}\,.
\end{equation}

\paragraph{Generalizations: }
Due to the existence of a consistent truncation of type IIB supergravity on any five-dimensional Einstein--Sasaki manifold $(X^5)$ to minimal $D=5$ gauged supergravity \cite{buchelliu}, we can extend the previous scaling to non-extremal R-charged black hole in $AdS_5\times X^5$.
Indeed, the metric for these black holes is given by \cite{kim1}:
\begin{equation}
  ds^2 = -\frac{1}{4}\,H^{-2}\,f\,dt^2 + H\left[f^{-1}\,dr^2 + r^2\,ds^2_{S^3}\right] + \left(d\chi + A\right)^2 + ds^2(KE^+_4)\,,
\end{equation}
where
\begin{eqnarray}
  H &=& 1 + \frac{q}{r^2}\,, \\
  f &=& 1 + r^2\,H^3 - \frac{\mu}{r^2}\,, \\
  A &=& \frac{1}{2}\,H^{-1}\,dt\,.
\end{eqnarray}
The scaling in this case works essentially as above.
The only subtlety lies on the focusing limit in the four-dimensional K\"ahler--Einstein manifold of positive curvature.
The giant gravitons are wrapping homologically trivial three-cycles there, and their location will be described, at least locally, by an extra coordinate.
Given the geometry of this base manifold, we could be more explicit, but we know the scaling has to satisfy
\begin{equation}
  ds^2(KE^+_4)\to \epsilon^2\,ds^2(\mathcal{M}_4)\,.
\end{equation}
Comparing to the detailed scaling in the three-charge case, the most obvious guess for $\mathcal{M}_4$ is that it will be a four-torus or a quotient thereof.
One way to think of this is that $\mathcal{M}_4$ is obtained from $KE^+_4$ by picking a point $P \in  KE^+_4$, to introduce Riemann normal coordinates $y^i$ around $P$ and to send $y^i \to \epsilon \, y^i$.
The metric on $KE^+_4$ is then reduced to $ds^2=g_{ij}(P) dy^i dy^j$, which is flat, and the only information about $KE^+_4$ that is retained is the precise range of values of the $y^i$.

\section{Discussion}

In this work we have studied the entropy of near-extremal R-charged $\ads{5}$ black holes.
Any attempt to count this entropy requires an identification of the microscopic degrees of freedom.
As was emphasized in \cite{superstar}, these R-charged black holes can be thought of as being built out of distributions of giant gravitons.
Since the latter are described in terms of spherical D3-branes, it is a natural idea that the microscopic degrees of freedom of non-extremal black holes are provided by the open strings stretched between the various giant gravitons \cite{gubserheckman,BHNL,davidemergence,Berensteinetal,BBFH,MelloKoch}.

Evidence in favor of this identification was provided at different levels:
\begin{itemize}
  \item[(i.)]
Using $\CN=4$ super-Yang--Mills, near-extremal black holes can be described as a gas of defects on top of a BPS condensate.
The counting of all such operators with the right quantum numbers is a combinatorial problem which we studied in this paper;
the results match a detailed computation of the gauge theory partition function at zero coupling.
Not surprisingly, we do not reproduce the detailed form of the entropy as obtained from the gravitational description, except for large non-extremality, a result that is well-known for neutral $\ads{5}$ black holes.
  \item[(ii.)]
The near-extremal gravitational entropy in the two- and three-charge systems is given by a function of the number of intersections between giant gravitons of different species.
This is very analogous to four-dimensional and five-dimensional black hole entropy countings in asymptotically (locally) flat spacetime, suggesting that there should be a microscopic interpretation in terms of the excitations of the open strings attached to these giants.
\end{itemize}

There is clearly a lot of work and understanding to be achieved in the gauge theory.
Any attempt to justify the gravitational entropy of these black holes will require a dynamical analysis in which coupling effects need to be understood.
This includes developing new and expanding existing technology to compute the anomalous dimensions of heavy operators $(\Delta \sim N^2)$, dealing with their mixing and potentially resumming infinite numbers of diagrams.
This last condition has to do with the fact that for these operators, {\em a priori}, non-planar diagrams cannot be neglected, and one may suspect that the diagram resummation may give rise to a new expansion parameter in $\CN=4$ super-Yang--Mills, similarly to what happened in the BMN sector \cite{BMN}.
Even the computations presented here did not appropriately include the information about the total number of giants.
In other words, there can be modifications to our zero coupling counting results from the analysis of ensembles in which the constraint on the number of giants (which is not a local conserved charge in the gauge theory) is properly considered.

The strongest confirmation of the physical picture advocated here comes from a careful study of the spacetime geometry close to the giant gravitons responsible for the BPS condensate.
These are distributed over the entire transverse five-sphere at locations parameterized by a two-sphere $\{\theta_1\,,\theta_2\}$ and generically rotating at the speed of light in the remaining angular directions.
The decoupling limits $(\ell_s\to 0)$ we found focused on the spacetime geometry close to the vicinity of a point in the two-sphere (a set of giants), while keeping the energy density of the excitations carried by these giants (the amount of non-extremality) properly fixed.
The latter is responsible for preserving the information about the existence of a horizon after taking the decoupling limit.
The entropy associated to these horizons follows from the usual Bekenstein--Hawking relation, and when integrated over the full set of allowed locations (two-sphere), reproduces the original $\ads{5}$ entropy in the near-extremal regime.

In the one-charge case, the decoupled geometries showed qualitatively the right behavior that one would expect for the near-horizon geometry of a stack of giant gravitons.
We also found that the decoupled geometry is invariant under an exchange of the original number of D3-branes and the number of giant gravitons, confirming a proposal in \cite{vijayasad}.

We should emphasize that we have not established the precise meaning of the families of CFTs that we found in the two-charge case.
Each of the CFTs carries a fraction of the total number of degrees of freedom and has central charge $c_{\text{strip}} = 6\,\delta N_2\,\delta N_3$, and adding all these contributions yields the right entropy.
However, we do not know how to compute the correlation functions of operators that live in separate CFTs.
It would be more appealing to have a single geometry describing all the degrees of freedom at once.
We have not found such a single geometry but discussed an interesting attempt in appendix~\ref{truesolution}.

In the two-charge system, the decoupled spacetimes describe a family of locally $\ads{3}$ geometries.
This family includes global $\ads{3}$ when the original $\ads{5}$ has vanishing $\mu$, conical defects when $\mu<\mu_c$, the massless non-rotating BTZ black holes when $\mu=\mu_c$, and massive non-rotating BTZ black holes when there is a non-vanishing horizon in $\ads{5}$.
This provides a satisfactory two-dimensional CFT explanation of the rather puzzling behavior of the two-charge black hole as a function of $\mu$.
Recall that for $\mu<\mu_c$ there is a naked singularity, and that only for $\mu>\mu_c$ a horizon forms.
In particular, the naked singularity is replaced in the decoupled geometry by a conical defect, suggesting that these geometries are all resolved in string theory.
In the context of D1-D5 physics, the $\ads{3}$ near-horizon geometry has been capped off by ``microstate-like'' solutions (see {\em e.g.}, \cite{rev1,rev2}).
It would be interesting to know whether similar constructions exist in this case.
Locally they do, but what would really be interesting is to find the global resolution in terms of the complete distribution of giants in the original $\ads{5}$.

In the three-charge system, we obtain $\ads{2}$ geometries, and so we suspect the dual description should be based on (super)conformal quantum mechanics, up to the usual problems associated to $\ads{2}$/CFT${}_1$ dualities.
The existence of these $\ads{2}$ geometries can be generalized to non-extremal R-charged black holes in $\AdS{5}\times X^5$, for any five-dimensional Einstein--Sasaki manifold $X^5$.
We do emphasize, however, that the entropy of these black holes is controlled by the square root of the number of pointlike giant intersections.
This could be interpreted as the number of Ramond--Ramond ground states of a string with central charge $c=N_1\,N_2\,N_3$.

\paragraph{Acknowledgments: }
We thank Ofer Aharony, Micha Berkooz, Jerome Gauntlett, Eric Gimon, Veronika Hubeny, Prem Kumar, Oleg Lunin, Samir Mathur, Asad Naqvi, Carlos Nu\~nez, Mukund Rangamani, Kyriakos Papadodimas, Simon Ross, and Samuel Vazquez for helpful discussions.
We thank the organizers of the Sowers Workshop in Theoretical Physics and the organizers of the Granada Workshop on Gravitational Aspects of String Theory during which some of this work was done.
V.B.\ thanks the theoretical physics group at UBC for hospitality while the paper was completed.
V.J.\ thanks the Benasque Center for Science for hospitality.
J.S.\ thanks the Theory Division at CERN for hospitality while this paper was completed.
V.B.\ was supported by the DOE under grant DE-FG02-95ER40893 and the NSF under grant OISE-0443607.
The research of JdB is supported financially by the Foundation of Fundamental Research on Matter (FOM).
V.J.\ is supported by PPARC.
J.S.\ was supported in part by DOE grant DE-AC02-05CH11231 and NSF grant PHY-0098840.

\appendix

\section{Half-BPS partition function rederivation}

\label{halfbps}

In this appendix we review how one explicitly derives the exact half-BPS partition function starting from the exact expression (\ref{genfie}):
\bea
Z & = & {\rm Tr}(x^{\Delta} q_1^{J_1} q_2^{J_2} q_3^{J_3})\nonumber \\
& = & \int [dU] \exp \left\{ \sum_{m=1}^{\infty} \frac{1}{m} (z_B(x^m,q_1^m,q_2^m,q_3^m) + (-1)^{m+1} z_F(x^m,q_1^m,q_2^m,q_3^m)) {\rm tr}(U^m) {\rm tr}(U^{\dagger})^m \right\} \,. \nonumber \\
& & \label{genfie11}
\eea
Note that ${\rm tr}(U) {\rm tr}(U^{\dagger})$ is the character of $U$, a unitary $N\times N$ matrix, in the adjoint representation.
Expanding the exponent yields a product of characters, and doing the group integral picks up the number of singlets, as required for physical operators.

The sum in the exponent arises because if we use more then one of the same building block in making an operator, we should (anti)symmetrize it appropriately, in order to count properly.
One can check that this is precisely what the sum with signs in the exponent accomplishes.
Alternatively, one can check that this type of alternating sum is exactly what reproduces the \Polya\ formula for the counting of the number of necklaces with given numbers and types of beads.

As an example, we pick out the terms which are half-BPS with respect to a given $U(1)$, {\em e.g.}, the one whose charge is measured by $q_1$.
In other words, we only want the terms that contain an equal power of $x$ and $q_1$.
Staring at $z_B$ and $z_F$ we see that the only term with this property is the term $xq_1$ in $z_B$.
Thus, the number of half-BPS states is counted by
\be \label{j1}
Z=\int [dU] \exp \left\{ \sum_{m=1}^{\infty} \frac{1}{m} x^m q_1^m {\rm tr}(U^m) {\rm tr}(U^{\dagger})^m \right\} \,.
\ee
This integral should be the same as the exact half-BPS partition function
\be
\prod_{k=1}^N \frac{1}{1-x^k q_1^k}
\ee
which one easily extracts from the free fermion representation.
Here we will show how one extracts this result from (\ref{j1}), using a few convenient equations collected from \cite{aux}.
First, we define
\be \label{eq21} \Delta(e^{i\theta}) = \prod_{k>m} (e^{i\theta_k}-e^{i\theta_m})
\ee
so that the integral over the unitary group, after passing to the eigenvalues, becomes
\be \label{eq22}
\int dU = \frac{1}{N!} \int_{0}^{2\pi} \ldots \int_0^{2\pi} \prod_i \frac{d\theta_i}{2\pi} \left| \Delta(e^{i\theta}) \right|^2 \,.
\ee
The character of a representation $R$ with highest weight components $m_i$, $i=1,\ldots,N$ is
\be \chi_R(e^{i\theta}) = \frac{\det_{km} (e^{il_k \theta_m})}{\Delta(e^{i\theta})} \,,
\label{char}
\ee
where the integers $l_k$ obey
\be l_1>l_2>l_3> \ldots > l_N
\ee
and are related to the highest weight component $m_k$ through
\be
l_k = m_k + N-k \,.
\ee
Of course the character (\ref{char}) is nothing but the Schur polynomial in the variables $x_k=e^{i\theta_k}$ for the partition $\{m_k\}$ of non-decreasing integers.
The Cauchy identity states that
\be \label{cauchy}
\prod_{k,l=1}^N \frac{1}{1-x_i y_j} = \sum_R' \chi_R(x) \chi_R(y) \,,
\ee
where the sum is over all representations $R$ for which the highest weights are non-negative, or in other words
\be m_1\geq m_2 \ldots \geq m_N \geq 0 \,, \qquad
l_1>l_2>l_3> \ldots > l_N \geq 0 \,.
\ee

With all this material, we return to (\ref{j1}).
Summing the exponent it can be rewritten as
\be
Z = \int [dU] {\rm \det}_{\rm adj}( (1 - x q_1 U)^{-1})
\ee
where the determinant is in the adjoint representation.
Next we pass to eigenvalues to write this as
\be \label{eq23}
Z= \int [dU] \prod_{i,j} (1- x q_1 e^{i(\theta_i-\theta_j)})^{-1} \,.
\ee
We can apply the Cauchy theorem (\ref{cauchy}) to this to recast it as
\be
Z = \int [dU] \sum_R' \chi_R(x q_1 e^{i\theta}) \chi_R(e^{i\theta})^{\ast} \,.
\ee
Now notice that
\be
\chi_R(\lambda x ) = \lambda^{\sum m_i} \chi_R(x)
\ee
if $R$ has highest weights $\{ m_k \}$.
Also, since characters are normalized with respect to the group measure, the integral of $\chi_R \chi_R^{\ast}$ is one for all $R$.
Therefore
\be
Z = \sum_{m_1\geq m_2 \ldots \geq m_N \geq 0} (x q_1)^{\sum m_i} = \prod_{k=1}^N \frac{1}{1-x^k q_1^k} \,,
\ee
which reproduces the correct answer.

This also provides a clue to to insert the giant graviton number into the game.
In applying the Cauchy identity, we need to restrict the sum over representations to highest weights with a fixed $m_1$.
It is not yet clear how to put this constraint directly into the full generating function (\ref{genfie11}).

\subsection{The half-BPS partition function approximation}

As mentioned above, the giant graviton constraint is not easy to implement in the exact partition function computation.
This is relevant because the non-extremal black hole has the same number of giants as the extremal one.
This means that any estimation of the partition function using some approximation may not take this into account.
It is natural to compute the entropy in the extremal single R-charge case and compare it to the one we computed in \cite{babel1}.

It is easy to derive the result:
\begin{equation}
  S \sim \frac{\pi\sqrt{2}}{\sqrt{3}}\,\sqrt{\Delta}\,.
\end{equation}
Thus, our method of extracting the dominant contribution to the partition function without the giant graviton constraint is able to capture the scaling with $N$, not surprisingly, since this was already the case for the hyperstar \cite{babel1}, but clearly misses all the functional dependence on the quotient $q_1/L^2=N_1/N$.

\section{Open string analysis}

A completely different approach to counting states is to work directly in terms of open strings stretched between giant gravitons. It would be interesting to explore this further but here we will just follow a totally na\"{\i}ve approach:
imagine $N$ branes and assume that there are $n_{i,j}$ oriented open strings starting at brane $i$ and ending at brane $j$.
To have a gauge invariant state, we need an equal number of open strings starting and ending on each brane.
That means that we have to impose the condition
\be \sum_{j} n_{i,j} = \sum_j n_{j,i}
\ee
for each $i$.
To impose this condition, we use a Lagrange multiplier $\theta_i$.
Since the condition is for integers, the Lagrange multiplier is an angular variable.
The total number of open string configurations, weighted with $q^{\sum_{i,j} n_{i,j}}$, becomes
\be
\label{eq11}
\prod_i \left( \int_{0}^{2\pi} \frac{d\theta_i}{2\pi} \right)  \sum_{n_{i,j}\geq 0} q^{\sum_{i,j} n_{i,j}}
\exp\left(i \sum_{i,j} \theta_i (n_{i,j} - n_{j,i})\right) \,.
\ee
This is equal to
\be
\label{eq12}
\prod_i \left( \int_{0}^{2\pi} \frac{d\theta_i}{2\pi} \right)  \sum_{n_{i,j}\geq 0}
\exp\left(i \sum_{i,j} n_{i,j} (\theta_i - \theta_j) \right) q^{\sum_{i,j} n_{i,j}} \,.
\ee
We can now do the sum over all integers $n_{i,j}$ and obtain
\be
\label{eq13}
\prod_i \left( \int_{0}^{2\pi} \frac{d\theta_i}{2\pi} \right) \prod_{i,j}
\frac{1}{1- q e^{i (\theta_i - \theta_j)}} \,.
\ee
This is almost the same as we get for the exact counting of half-BPS states (see (\ref{eq23})).
The only difference is that the measure is different (see (\ref{eq22})).
This is probably due to the fact that we have not imposed the permutation symmetry between the branes.
We have not checked this explicitly, but one would expect that imposing invariance under permutations of the branes will precisely yield the measure of $U(N)$ --- it is hard to see how anything else could come out.
The above result is then the counting of the number of branes for separated, distinguishable branes.
It would be interesting to see if this can be used for a counting of the number of states of R-charged black holes, and also to generalize it to open strings stretched between (dual) giants.

\section{A new solution of type IIB supergravity}

\label{truesolution}

As explained at the end of section~(\ref{2charge}), it would be interesting to find a metric containing global information about the distribution of giant gravitons in the original $AdS_5$ geometry.
The most na\"{\i}ve metric candidate is (\ref{2scalemet1}) with $\mu_i^0$ replaced by $\mu_i$.
Such a replacement is achieved if one does not focus on a given $S^2$ point $\{\mu_i^0\}$, {\em i.e.}, if one does not scale the $\theta_i$ in  (\ref{sca1})--(\ref{sca4}). If we also drop the rescalings of $\chi_{2,3}$, we find that the metric  schematically scales as
\begin{equation}
  ds^2 = L^2 \cos\theta_1 (\epsilon^2 (ds_6^2 + \ldots) + (ds_4^2 + \ldots)) \,, \label{aux2}
\end{equation}
where $ds_6^2$ stands for the $AdS_3$ and $S^3$ factors, whereas $ds_4^2$ describes the remaining four dimensions in the original $S^5$ not belonging to the circle where the two distributions of giants intersect. This metric does not scale homogeneously, and so one can not use analyticity to argue that it solves the type IIB equations of motion. We have indeed checked that it does not, when including the corresponding RR 5-form fluxes.

\paragraph{A new solution of type IIB supergravity: }

We report here on a new solution to type IIB equations of motion. This is obtained by setting $\epsilon$ to one in the above metric
\begin{multline}
ds^2 = \mu_1 \left[ - {r^2 \over \sqrt{q_2 q_3}} \, f \, dt^2 + {\sqrt{q_2 q_3} \over r^2}\frac{1}{f}
 \, dr^2 + {L^2 \, r^2 \over \sqrt{q_2 q_3}} \, d\phi_1^2 +  \sqrt{q_2 q_3} \, ds^2_{S^3} \right]  \\
+ {L^2 \over \mu_1 \sqrt{q_2 q_3}} \left[ \sum_{i=2,3} q_i (d\mu_i^2 + (\mu_i)^2 \, d\chi_i^2) \right] \,, \label{2scalemet1a}
\end{multline}
with
\be \label{hup}
f = 1 - \frac{\mu}{r^2} + \frac{q_2+q_3}{L^2}  + {q_2 q_3 \over L^2 r^2} \,.
\ee
rescaling the five-form by a factor of
\be
F^{(5)} \rightarrow \left( 1 + \frac{{q}_2 + {q}_3}{2L^2} \right) F^{(5)}.
\ee
and finally, setting ${q}_2={q}_3$.

This solution describes a warped $\ads{3}$ metric and does contain global information about the distribution of giant gravitons. We have not been able to match precisely our metric with the large class of $1/8$-supersymmetric warped $\AdS{3}$ metrics in type IIB supergravity studied in \cite{kim1,kim2}  or to the $1/4$-BPS solutions described in \cite{donos}.   While we do not understand the origin of this solution, the fact that it involves a small modification of the original metric for ${q}_{2,3}\ll L^2$, which was the same regime in which the entropies agreed, suggests that perhaps there is some systematic way to generate such exact solutions as an expansion in $\frac{q_2}{L^2},\frac{q_3}{L^2}$.

Schematically, the exact solution looks like
\be
L^{-2} ds^2 = \sqrt{1-\rho^2}\,(ds^2_{\ads{3}} + ds^2_{S^3}) + \frac{1}{\sqrt{1-\rho^2}}\,(d\rho^2 + \rho^2 ds^2_{S'_3})
\ee
and one can check that such solutions only exist if the curvature radii of $\ads{3}$ and the $S^3$ are different.
This metric is reminiscent of that of a brane wrapping $\ads{3}\times S^3$, albeit with the usual harmonic functions replaced by $1-\rho^2$, with $\rho$ the radial coordinate on $\mathbb R^4$.
We leave a further exploration of these types of metrics to future work.

\newpage

\providecommand{\href}[2]{#2}\begingroup\raggedright

\endgroup


\begin{thebibliography}{10}

\bbibitem{cveticetal}
 M.~Cvetic {\it et al.},
  ``Embedding AdS black holes in ten and eleven dimensions,''
  Nucl.\ Phys.\  B {\bf 558}, 96 (1999)
  [arXiv:hep-th/9903214].

 \bbibitem{superstar}
  R.~C.~Myers and O.~Tafjord,
  ``Superstars and giant gravitons,''
  JHEP {\bf 0111}, 009 (2001)
  [arXiv:hep-th/0109127].



\bbibitem{babel1}
  V.~Balasubramanian, V.~Jejjala, and J.~Sim\'on,
  ``The library of Babel,''
  Int.\ J.\ Mod.\ Phys.\ D {\bf 14}, 2181 (2005)
  [arXiv:hep-th/0505123].

  V.~Balasubramanian, J.~de Boer, V.~Jejjala, and J.~Sim\'on,
  ``The library of Babel: On the origin of gravitational thermodynamics,''
  JHEP {\bf 0512}, 006 (2005)
  [arXiv:hep-th/0508023].

\bbibitem{quantgrav}
  V.~Balasubramanian, B.~Czech, K.~Larjo, D.~Marolf and J.~Simon,
  ``Quantum geometry and gravitational entropy,''
  arXiv:0705.4431 [hep-th].





\bbibitem{BBNS} V.~Balasubramanian, M.~Berkooz, A.~Naqvi and
M.~J.~Strassler,
  ``Giant gravitons in conformal field theory,''
  JHEP {\bf 0204}, 034 (2002)
  [arXiv:hep-th/0107119].

\bbibitem{cjr}
 S.~Corley, A.~Jevicki and S.~Ramgoolam,
  ``Exact correlators of giant gravitons from dual N = 4 SYM theory,''
  Adv.\ Theor.\ Math.\ Phys.\  {\bf 5}, 809 (2002)
  [arXiv:hep-th/0111222].

\bbibitem{ber}
 D.~Berenstein,
  ``A toy model for the AdS/CFT correspondence,''
  JHEP {\bf 0407}, 018 (2004)
  [arXiv:hep-th/0403110].



\bbibitem{BHNL}
  V.~Balasubramanian, M.~x.~Huang, T.~S.~Levi and A.~Naqvi,
  ``Open strings from N = 4 super Yang-Mills,''
  JHEP {\bf 0208}, 037 (2002)
  [arXiv:hep-th/0204196].

\bbibitem{davidemergence}
 D.~Berenstein,
  ``Large N BPS states and emergent quantum gravity,''
  JHEP {\bf 0601}, 125 (2006)
  [arXiv:hep-th/0507203].

\bbibitem{Berensteinetal}
  D.~Berenstein and S.~E.~Vazquez,
  ``Integrable open spin chains from giant gravitons,''
  JHEP {\bf 0506}, 059 (2005)
  [arXiv:hep-th/0501078].

   D.~Berenstein, D.~H.~Correa and S.~E.~Vazquez,
  ``Quantizing open spin chains with variable length: An example from giant
  gravitons,''
  Phys.\ Rev.\ Lett.\  {\bf 95}, 191601 (2005)
  [arXiv:hep-th/0502172].

    D.~Berenstein, D.~H.~Correa and S.~E.~Vazquez,
  ``A study of open strings ending on giant gravitons, spin chains and
  integrability,''
  JHEP {\bf 0609}, 065 (2006)
  [arXiv:hep-th/0604123].



\bbibitem{BBFH}
  V.~Balasubramanian, D.~Berenstein, B.~Feng and M.~x.~Huang,
  ``D-branes in Yang-Mills theory and emergent gauge symmetry,''
  JHEP {\bf 0503}, 006 (2005)
  [arXiv:hep-th/0411205].



\bbibitem{MelloKoch}
 R.~de Mello Koch and R.~Gwyn,
  ``Giant graviton correlators from dual SU(N) super Yang-Mills theory,''
  JHEP {\bf 0411}, 081 (2004)
  [arXiv:hep-th/0410236].

  R.~de Mello Koch, J.~Smolic and M.~Smolic,
  ``Giant Gravitons - with Strings Attached (I),''
  arXiv:hep-th/0701066.

    R.~de Mello Koch, J.~Smolic and M.~Smolic,
  ``Giant Gravitons - with Strings Attached (II),''
  arXiv:hep-th/0701067.


\bbibitem{sonner}
A.~Sinha and J.~Sonner,
  ``Black Hole Giants,''
  arXiv:0705.0373 [hep-th].

\bbibitem{vijayasad}
  V.~Balasubramanian and A.~Naqvi,
  ``Giant gravitons and a correspondence principle,''
  Phys.\ Lett.\  B {\bf 528}, 111 (2002)
  [arXiv:hep-th/0111163].

  \bbibitem{gubserheckman}
  S.~S.~Gubser and J.~J.~Heckman,
  ``Thermodynamics of R-charged black holes in AdS(5) from effective
  strings,''
  JHEP {\bf 0411}, 052 (2004)
  [arXiv:hep-th/0411001].


\bbibitem{llm}
  H.~Lin, O.~Lunin, and J.~M.~Maldacena,
  ``Bubbling AdS space and 1/2 BPS geometries,''
  JHEP {\bf 0410}, 025 (2004)
  [arXiv:hep-th/0409174].

\bbibitem{BDHM}
 T.~Banks, M.~R.~Douglas, G.~T.~Horowitz and E.~J.~Martinec,
  ``AdS dynamics from conformal field theory,''
  arXiv:hep-th/9808016.


\bbibitem{cveticgubser}
  M.~Cvetic and S.~S.~Gubser,
   ``Phases of R-charged black holes, spinning branes and strongly coupled gauge theories,''
  JHEP {\bf 9904}, 024 (1999)
  [arXiv:hep-th/9902195].

\bbibitem{cejm}
  A.~Chamblin, R.~Emparan, C.~V.~Johnson and R.~C.~Myers,
  ``Charged AdS black holes and catastrophic holography,''
  Phys.\ Rev.\  D {\bf 60}, 064018 (1999)
  [arXiv:hep-th/9902170].

  A.~Chamblin, R.~Emparan, C.~V.~Johnson and R.~C.~Myers,
  ``Holography, thermodynamics and fluctuations of charged AdS black holes,''
  Phys.\ Rev.\  D {\bf 60}, 104026 (1999)
  [arXiv:hep-th/9904197].

\bbibitem{gubsermitra}
  S.~S.~Gubser and I.~Mitra,
  ``Instability of charged black holes in anti-de Sitter space,''
  arXiv:hep-th/0009126.

\bbibitem{gubsermitra1}
  S.~S.~Gubser and I.~Mitra,
  ``The evolution of unstable black holes in anti-de Sitter space,''
  JHEP {\bf 0108}, 018 (2001)
  [arXiv:hep-th/0011127].


\bbibitem{yaya}
  D.~Yamada and L.~G.~Yaffe,
  ``Phase diagram of N = 4 super-Yang-Mills theory with R-symmetry chemical
  potentials,''
  JHEP {\bf 0609}, 027 (2006)
  [arXiv:hep-th/0602074].



\bibitem{othershalfBPS}
  A.~Buchel,
  ``Coarse-graining 1/2 BPS geometries of type IIB supergravity,''
  Int.\ J.\ Mod.\ Phys.\  A {\bf 21}, 3495 (2006)
  [arXiv:hep-th/0409271].

  N.~V.~Suryanarayana,
  ``Half-BPS giants, free fermions and microstates of superstars,''
  JHEP {\bf 0601}, 082 (2006)
  [arXiv:hep-th/0411145].

  P.~G.~Shepard,
  ``Black hole statistics from holography,''
  JHEP {\bf 0510}, 072 (2005)
  [arXiv:hep-th/0507260].

    P.~J.~Silva,
  ``Rational foundation of GR in terms of statistical mechanic in the  AdS/CFT
  framework,''
  JHEP {\bf 0511}, 012 (2005)
  [arXiv:hep-th/0508081].




\bbibitem{harvard}
  O.~Aharony, J.~Marsano, S.~Minwalla, K.~Papadodimas and M.~Van Raamsdonk,
  ``The Hagedorn / deconfinement phase transition in weakly coupled large N
  gauge theories,''
  Adv.\ Theor.\ Math.\ Phys.\  {\bf 8}, 603 (2004)
  [arXiv:hep-th/0310285].

\bibitem{harmark}
T.~Harmark and M.~Orselli,
  ``Quantum mechanical sectors in thermal N = 4 super Yang-Mills on R x
  S**3,''
  Nucl.\ Phys.\  B {\bf 757}, 117 (2006)
  [arXiv:hep-th/0605234].

\bbibitem{hopo}
  G.~T.~Horowitz and J.~Polchinski,
  ``A correspondence principle for black holes and strings,''
  Phys.\ Rev.\  D {\bf 55}, 6189 (1997)
  [arXiv:hep-th/9612146].



  \bbibitem{juanAdS}
  J.~M.~Maldacena,
  ``The large N limit of superconformal field theories and supergravity,''
  Adv.\ Theor.\ Math.\ Phys.\  {\bf 2}, 231 (1998)
  [Int.\ J.\ Theor.\ Phys.\  {\bf 38}, 1113 (1999)]
  [arXiv:hep-th/9711200].



\bbibitem{MST}
  J.~McGreevy, L.~Susskind and N.~Toumbas,
  ``Invasion of the giant gravitons from anti-de Sitter space,''
  JHEP {\bf 0006}, 008 (2000)
  [arXiv:hep-th/0003075].

\bibitem{shahin}
M.~M.~Sheikh-Jabbari, ``Tiny graviton matrix theory: DLCQ of IIB
plane-wave string theory, a conjecture,'' JHEP {\bf 0409}, 017
(2004) [arXiv:hep-th/0406214].


\bbibitem{IH}
    A.~Hashimoto, S.~Hirano and N.~Itzhaki,
  ``Large branes in AdS and their field theory dual,''
  JHEP {\bf 0008}, 051 (2000)
  [arXiv:hep-th/0008016].


\bbibitem{oldnote}
V.~Balasubramanian, M.-x.~Huang, T.~Levi, A.~Naqvi, ``A Geometric Correspondence'', unpublished notes.


\bibitem{iks}
N.~Itzhaki, D.~Kutasov and N.~Seiberg, ``I-brane dynamics,'' JHEP
{\bf 0601}, 119 (2006) [arXiv:hep-th/0508025].


\bibitem{ms}
A.~E.~Mosaffa and M.~M.~Sheikh-Jabbari, ``On classification of the
bubbling geometries,'' JHEP {\bf 0604}, 045 (2006)
[arXiv:hep-th/0602270].


\bibitem{kim1}
  J.~P.~Gauntlett, N.~Kim and D.~Waldram,
  ``Supersymmetric AdS(3), AdS(2) and bubble solutions,''
  JHEP {\bf 0704}, 005 (2007)
  [arXiv:hep-th/0612253].


\bibitem{kim2}
  N.~Kim,
  ``AdS(3) solutions of IIB supergravity from D3-branes,''
  JHEP {\bf 0601}, 094 (2006)
  [arXiv:hep-th/0511029].

\bbibitem{conical}
  V.~Balasubramanian, J.~de Boer, E.~Keski-Vakkuri and S.~F.~Ross,
  ``Supersymmetric conical defects: Towards a string theoretic description  of
  black hole formation,''
  Phys.\ Rev.\  D {\bf 64}, 064011 (2001)
  [arXiv:hep-th/0011217].

\bbibitem{maldmaoz}
  J.~M.~Maldacena and L.~Maoz,
  ``De-singularization by rotation,''
  JHEP {\bf 0212}, 055 (2002)
  [arXiv:hep-th/0012025].

\bbibitem{mathurlunin}  O.~Lunin, S.~D.~Mathur and A.~Saxena,
  ``What is the gravity dual of a chiral primary?,''
  Nucl.\ Phys.\  B {\bf 655}, 185 (2003)
  [arXiv:hep-th/0211292].

\bbibitem{maldmaozlunin}
  O.~Lunin, J.~M.~Maldacena and L.~Maoz,
  ``Gravity solutions for the D1-D5 system with angular momentum,''
  arXiv:hep-th/0212210.

\bbibitem{adsentropies}
  M.~Banados,
  ``Global Charges In Chern-Simons Field Theory And The (2+1) Black Hole,''
  Phys.\ Rev.\  D {\bf 52}, 5816 (1996)
  [arXiv:hep-th/9405171].


\bibitem{mikhailov}
  A.~Mikhailov,
  ``Giant gravitons from holomorphic surfaces,''
  JHEP {\bf 0011}, 027 (2000)
  [arXiv:hep-th/0010206].

\bibitem{stromvafa}
  A.~Strominger and C.~Vafa,
  ``Microscopic Origin of the Bekenstein-Hawking Entropy,''
  Phys.\ Lett.\  B {\bf 379}, 99 (1996)
  [arXiv:hep-th/9601029].



\bibitem{buchelliu}
  A.~Buchel and J.~T.~Liu,
  ``Gauged supergravity from type IIB string theory on Y(p,q) manifolds,''
  Nucl.\ Phys.\  B {\bf 771}, 93 (2007)
  [arXiv:hep-th/0608002].

 \bbibitem{BMN}
  D.~Berenstein, J.~M.~Maldacena and H.~Nastase,
  ``Strings in flat space and pp waves from N = 4 super Yang Mills,''
  JHEP {\bf 0204}, 013 (2002)
  [arXiv:hep-th/0202021];
  N.~R.~Constable, D.~Z.~Freedman, M.~Headrick, S.~Minwalla, L.~Motl,
A.~Postnikov and W.~Skiba,
  ``PP-wave string interactions from perturbative Yang-Mills theory,''
  JHEP {\bf 0207}, 017 (2002)
  [arXiv:hep-th/0205089].




\bibitem{rev1}
S.~D.~Mathur,
  ``The fuzzball proposal for black holes: An elementary review,''
  Fortsch.\ Phys.\  {\bf 53}, 793 (2005)
  [arXiv:hep-th/0502050].

\bibitem{rev2}
I.~Bena and N.~P.~Warner,
  ``Black holes, black rings and their microstates,''
  arXiv:hep-th/0701216.

\bbibitem{aux}
D.~Boulatov and V.~Kazakov,
  ``One-Dimensional String Theory With Vortices As The Upside Down Matrix
  Oscillator,''
  Int.\ J.\ Mod.\ Phys.\  A {\bf 8}, 809 (1993)
  [arXiv:hep-th/0012228].

\bibitem{donos}  A.~Donos,
  ``A description of 1/4 BPS configurations in minimal type IIB SUGRA,''
  Phys.\ Rev.\  D {\bf 75}, 025010 (2007)
  [arXiv:hep-th/0606199].










\end{thebibliography}
\end{document}